\documentclass[acmtog,nonacm]{acmart}

\author{Arjun S. Lakshmipathy}
\affiliation{%
   \institution{Carnegie Mellon University, FAIR at Meta}%
   \country{USA}
}

\author{Jessica K. Hodgins}
\affiliation{%
   \institution{Carnegie Mellon University, Boston Dynamics AI Institute}%
   \country{USA}
}

\author{Nancy S. Pollard}
\affiliation{%
   \institution{Carnegie Mellon University}%
   \country{USA}
}

\acmSubmissionID{???}

\usepackage{booktabs} 

\usepackage[ruled]{algorithm2e} 
\usepackage{graphicx, caption, subcaption}
\usepackage{mathrsfs}
\usepackage{enumerate}
\graphicspath{ {figures/} }

\DeclareMathOperator*{\argmin}{arg\,min}

\newcommand{\vect}[1]{\boldsymbol{#1}}

\newcommand{\enumeratext}[1]{%
\setcounter{saveenumerate}{\value{enum\romannumeral\the\@enumdepth}}
\end{enumerate}
#1
\begin{enumerate}
\setcounter{enum\romannumeral\the\@enumdepth}{\value{saveenumerate}}%
}

\usepackage{array}
\newcolumntype{P}[1]{>{\centering\arraybackslash}p{#1}}
\newcolumntype{M}[1]{>{\centering\arraybackslash}m{#1}}

\SetAlFnt{\small}
\SetAlCapFnt{\small}
\SetAlCapNameFnt{\small}
\SetAlCapHSkip{0pt}

\copyrightyear{2024}
\acmYear{2024}
\acmMonth{1}

\begin{document}

\title{Kinematic Motion Retargeting for Contact-Rich Anthropomorphic Manipulations}

\begin{abstract}

Hand motion capture data is now relatively easy to obtain, even for complicated grasps; however this data is of limited use without the ability to retarget it onto the hands of a specific character or robot. The target hand may differ dramatically in geometry, number of degrees of freedom (DOFs), or number of fingers. We present a simple, but effective framework capable of kinematically retargeting multiple human hand-object manipulations from a publicly available dataset to a wide assortment of kinematically and morphologically diverse target hands through the exploitation of contact areas. We do so by formulating the retarget operation as a non-isometric shape matching problem and use a combination of both surface contact and marker data to progressively estimate, refine, and fit the final target hand trajectory using inverse kinematics (IK). Foundational to our framework is the introduction of a novel shape matching process, which we show enables predictable and robust transfer of contact data over full manipulations while providing an intuitive means for artists to specify correspondences with relatively few inputs. We validate our framework through thirty demonstrations across five different hand shapes and six motions of different objects. We additionally compare our method against existing hand retargeting approaches. Finally, we demonstrate our method enabling novel capabilities such as object substitution and the ability to visualize the impact of design choices over full trajectories.

\end{abstract}

%
%


%
%


\maketitle

\section{Introduction}

\begin{figure}
\begin{center}
\includegraphics[width=1.0\linewidth]{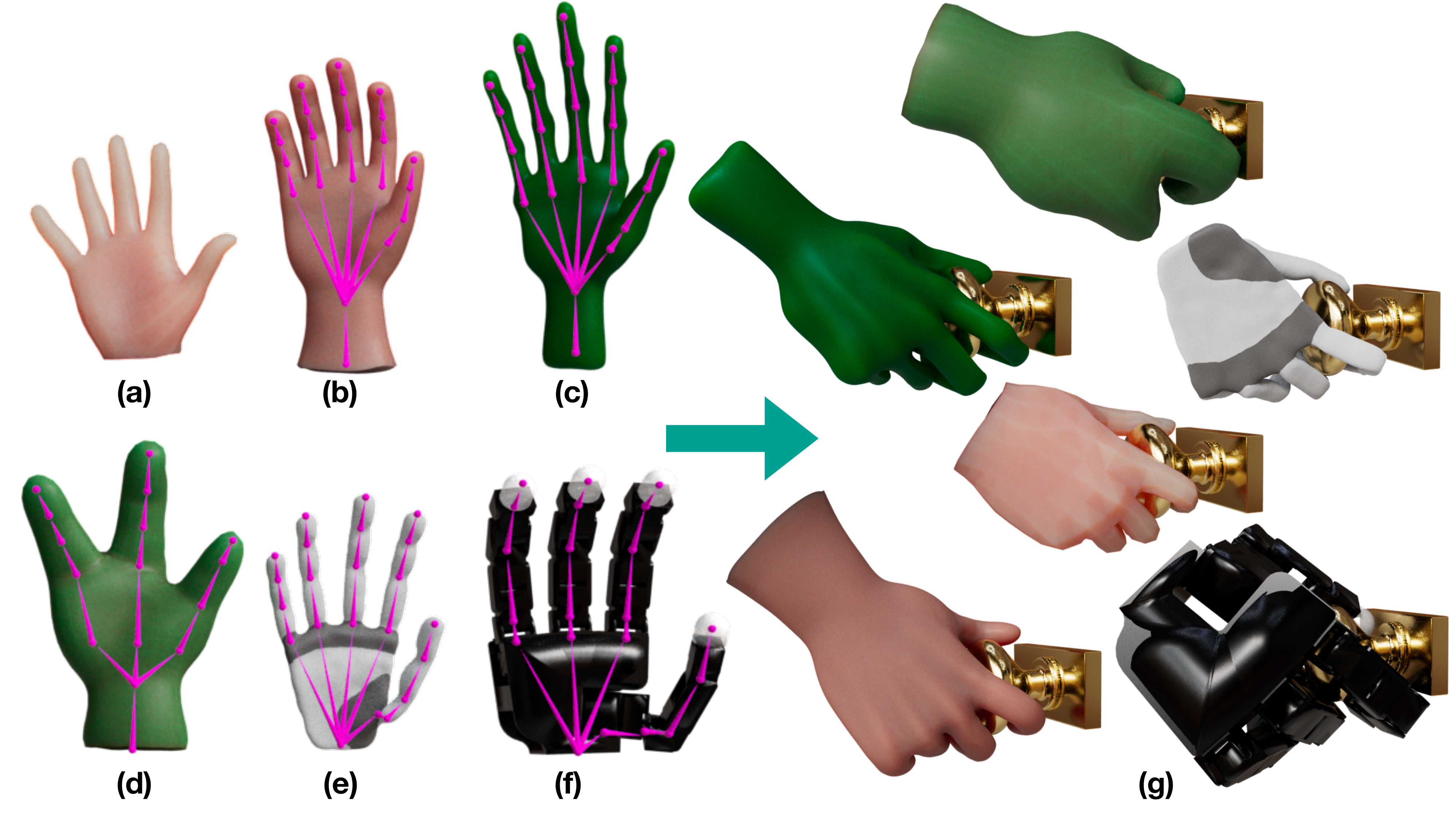}
\caption{All hands used in our experiments, including the (a) source MANO hand, (b) an alternate human hand, (c) a witch hand, (d) an alien hand, (e) a custom prosthetic hand, and (f) the Allegro Hand. (g) We retarget demonstrations performed by the source hand to all these hands by procedurally transferring contact areas over the entire time series via shape matching.}
\label{fig:allhands}
\end{center}
\end{figure}

Advancements in motion capture (mocap) technology have made it possible to collect motion data at high levels of detail, including large scale body movements and fine-grained hand movements together in a single capture \cite{romero2017embodied}; however, \textit{using} this data is still difficult because the targeted embodiment of the data is almost never the same as the demonstrator. For hands in particular, retargeting is often required because of the many different desired hand models and the precision required to make grasps, especially those with many points of contact, look compelling. Unfortunately, this process commonly generates artifacts such as motion misalignment or lack of contact that are difficult to repair in post-processing. The problem becomes even more challenging when the target hand is not human-like, whether that includes different finger proportions, DOFs, or number of fingers.

The lack of reliable retargeting methods for adapting existing contact-rich hand motion data --- specifically data that has been previously collected for a hand other than the desired target --- remains a consistent bottleneck for multiple communities. In film and game production, considerable animator time is spent customizing common manipulations to different character rigs. In robot learning, the difficulty in adapting existing manipulation data to different hands contributes to the data sparsity problem \cite{sivakumar2022telekinesis}, which in turn limits the feasibility of supervised and semi-supervised machine learning algorithms. A standardized approach for re-purposing existing data, and in particular data easily collectible from humans, would provide a much-needed solution for a broad range of applications.

This paper presents an approach for kinematic retargeting of existing contact-rich anthropomorphic hand-object manipulations through the use of dense correspondences (guaranteed bijections of discretized point sets) between contact areas. To do so, we treat time-varying hand contact distributions as a foundational retargeting medium and formulate the mapping of distributions between different hands as a non-isometric shape matching problem. We then show that exploiting retargeted contact correspondences between hands and objects leads to the development of a straightforward motion synthesis pipeline. Despite its simplicity, our approach is robust, predictable, and effective, which we demonstrate by mapping motions of varying complexity from the publicly available GRAB dataset \cite{taheri2020grab} to five kinematically, morphologically, and geometrically diverse hands sourced as-is from artist rigs intended for media production and robot manipulators intended for simulation.

Specifically, we make the following novel technical contributions:

\begin{itemize}
    \item A local shape matching algorithm for robust time-series contact transfer
    \item A multi-stage optimization pipeline to compute the retargeted motion
\end{itemize}

Although the optimization pipeline is straightforward, the proposed shape matching technique based on local charts enables obtaining dense correspondences in a satisfactory way based on an intuitive markup scheme. We then thoroughly evaluate our method with the following results:

\begin{itemize}
    \item Thirty retargeted demonstrations from a publicly available dataset (5 hands x 6 motions of different objects)
    \item Quantitative evaluation of inter-object, table, and self intersections 
    \item Novel extensions, including visualizing the impact of design parameters and object substitution
    \item Baseline comparisons to validate the importance of contact information
\end{itemize}

The techniques and algorithms proposed in this paper are designed for \textbf{standardization and simplicity,} which we show enable \textbf{reliable generation of high quality retargeted hand motion trajectories}. Our decision to focus on kinematics, rather than dynamics, in this work is primarily motivated by reliability and secondarily by speed. Despite this simplification, we show that our method produces high quality results, generalizes well across manipulators, and gracefully handles complex manipulation tasks with many and changing contacts. Depending on the application, it can be considered complete as-is or a first pass for additional processing.

\section{Related Work}

We consider three categories of relevant prior work: motion retargeting, shape matching, and contact-driven grasp synthesis.

\subsection{Motion Retargeting}

Adapting existing motion data to new characters is a long-standing and well-studied problem for both full-body and hand animation.

Joint-based retargeting remains a common approach in full-body character animation; however, resolving environmental and self-contact events complicates the process. Proposed solutions include adapting to variations in body shape using learning-based methods exclusively \cite{aberman2020skeleton} or augmented by physics-based simulation \cite{won2019bodyshapevariation, ryu2021muscleretarget}, differentiating between unwanted self-contacts and desired foot contacts during optimization \cite{villegas2021skinnedmotion}, exploiting spatial maps \cite{kim2016retargeting}, standardized wrapper meshes \cite{jin2018aura} or interaction graphs \cite{zhang2023multicharacter}, and dynamics-augmented projected motion optimization \cite{lee2019projective}. These methods have produced useful results; however, a key distinction between hand and full body retargeting is the number of contact points required to make a motion visually compelling. More concretely, single point contacts between objects and different body parts have proven largely sufficient at the full body scale even for transferring highly complex interactions between multiple characters and objects \cite{zhang2023multicharacter}. Conversely, a small number of points has been shown to insufficiently model the complexity of realistic interaction between hands and objects \cite{lakshmipathy2023contactedit}. Additionally, as illustrated in Figure~\ref{fig:allhands}, the problem is further compounded by the fact that hand shapes, morphologies, and kinematics can vary widely between different characters.

Common approaches to retargeting motions from hand tracking include direct joint mapping \cite{rajeswaran2018dapg,kumar2015haptix}, keypoint-based IK \cite{humbertson2015handson,qin2022from,antotsiou2018taskretargeting,dasari2023pgdm}, inverse dynamics from input joint angles \cite{kry2006interactioncapture}, and hand shape via relative vector distances (henceforth referred to as ``functional pose equivalence") \cite{Handa2019DexPilotVT,sivakumar2022telekinesis}. The latter methods remain state of the art approaches when the hands are more divergent. These approaches can also be adapted to previously collected data; however, we show that doing so often creates significant artifacts which are difficult to clean up in post-processing.

Physics simulation is also commonly used to improve visual plausibility and Sim2Real transfer success. Due to the complexities of frequent making and breaking of contacts during manipulation, several approaches employ learning from demonstration (LfD) to generate physics-based target hand policies from human motion data \cite{qin2022from,wu2022ILAD,dasari2023pgdm}. Standard kinematic retargeting techniques (e.g. direct joint mapping, keypoint or keyvector IK) are often used in these approaches to create expert trajectories and subsequently pre-populate reward tables; however, this strategy can result in failures or unexpected results if the retargeted hand trajectory is poor. Identifying retargeting failures, as well as disentangling error contributions from policy learning and reward shaping, is especially challenging at scale. Our goal in this work is to provide a solution for reliable upstream expert trajectory estimation that can be used in conjunction with any downstream physics simulator.

There is currently \textit{no standardized solution} for reliably retargeting existing hand-object motion data. We argue in this work that contact information is vital to produce high quality results across a broad range of motions and hands.

\subsection{Shape Matching}

Shape matching addresses the problem of finding geometrically meaningful correspondences between two different shapes. Existing literature in this space can be approximately divided into isometric and non-isometric problem domains. We provide a brief summary of each variant.

Isometric shape matching concerns instances where the baseline shape is (approximately) the same. Common examples include character meshes which have undergone skeleton-driven deformation or deformable objects. The key assumption in this domain is that geodesic distances, and equivalently the Laplace-Beltrami operator, are preserved. This assumption enables the use of methods such as functional maps with consistent descriptor functions \cite{ovsjanikov2012functional,pai2021sinkhorn,attaiki2021dpfm} and heat diffusion \cite{sharp2022diffusion} to identify equivalences. The reliance on distance preservation makes problems in this domain comparatively easier; unfortunately, these techniques are rarely applicable to our domain due to the large range of geometric variations in hand meshes.

Non-isometric shape matching relaxes the distance preservation assumption, thereby extending the problem space to fully differentiated shapes at the cost of substantially escalating problem difficulty. The majority of existing works in this space consider mapping of global media (e.g. textures) and require some form of user input to compensate for the lack of reliable quantitative measures of equivalence, typically in the form of landmark points \cite{panine2022landmarkadapted, ezuz2019reversibleharmonic} or curves \cite{gehre2018interactivecurve}. Landmarks are subsequently used as boundary conditions \cite{panine2022landmarkadapted,gehre2018interactivecurve}, as keypoints for the extraction of hyperbolic Tutte embeddings \cite{aigerman2016orbifold,takayama2022compatibletriangulations}, or other demarcations. Although we tried several such techniques, we found that they were not robust enough to handle all of the variations in our hand and object geometries and generally struggled with exact mappings of individual points (point-to-point transfers). We show in this work that relaxing the need to map globally smooth media enables the use of alternate strategies capable of performing precise point-to-point transfers from similar landmark annotations. Specifically, we formulate shape correspondence being governed by an atlas, where each constituent chart is a logarithmic map \cite{schmidt2006dem}.

Efforts have been made to fully automate the non-isometric case, including automating the discovery of landmarks though heuristics \cite{edelstein2020evolutionarynonisometric,marin2020farm}, aligning extrinsic local surface correspondences such as normals \cite{li2007shape}, and automatically segmenting meshes via data-driven techniques \cite{kalogerakis2010labelMeshes}. However, the immense variety in hand designs (number of finger segments, finger lengths, finger counts, palm shapes, etc.) compounded with their ability to change shape with pose render the application of heuristics or extrinsic metrics particularly challenging. Existing pose data for arbitrary hand models is also difficult to obtain as it is simply not possible to cover the entire space of all possible hand designs. Furthermore, fully automated methods typically permit little control over the results, which can render such techniques unusable for quality-critical applications. In this work, we instead focus on creating highly reliable mappings for \textit{direct}, local data transfer using simple curve-based annotations easily obtainable from an artist. We also note that our approach is \textit{dependent only on intrinsic quantities}, which inherently makes it robust to geometric variations in base design as well as deformation induced by pose changes.

\subsection{Contact-Driven Grasp Synthesis}

Contact points have played a vital role in the generation of grasps and manipulations. Techniques include targeting optimal independent regions within which point contacts can be placed \cite{roa2009icr3d} or using contacts as error terms in the training of generative models \cite{christen2021dgrasp, wu2021SAGA}. A number of approaches target manipulation planning and optimization by switching between contact modes \cite{cheng2021contactmode2d} or by optimizing single point contact placement and forces for physically plausible results \cite{mordatch2012contact,hazard2020automated,ye2012contactsampling}. The $\epsilon$ metric proposed by Ferrari and Canny \cite{ferrari1992metric} for evaluating grasp wrench spaces --- the most widely used grasp quality metric --- is fundamentally rooted in its analysis at individual contact points.

However, characterization of grasps as single points highly simplifies the complexities of real interactions \cite{brahmbhatt2019contactdb,brahmbhatt2020contactpose} and cannot account for geometric consistency outside the designated point. Consequently, there has been considerable recent interest in collecting, analyzing, and exploiting contact areas kinematically \cite{,brahmbhatt2019contactgrasp,lakshmipathy2021contacttracing,lakshmipathy2022contacttransfer,taheri2020grab,fan2023arctic} and dynamically \cite{turpin2022graspd,pang2022globalplanning}. We focus on kinematics in this paper, following recent precedents that generate and optimize poses by matching contact patches \cite{wei2023generalized, brahmbhatt2019contactgrasp,lakshmipathy2022contacttransfer,lakshmipathy2023contactedit,grady2021contactopt}. Our primary goal in this paper is to establish the techniques necessary to adapt contact patch-driven grasping strategies to full manipulation trajectories. Although this idea appears straightforward, it was not possible to do so without completely rethinking the problem of shape matching due to the dense, complex, and constantly changing collections of contact areas that had to be transferred meaningfully between different hands.

\section{Method}
\label{sec:methods}

\begin{figure*}
\centering
\includegraphics[trim={0 0 0 0},clip,width=1.0\linewidth]{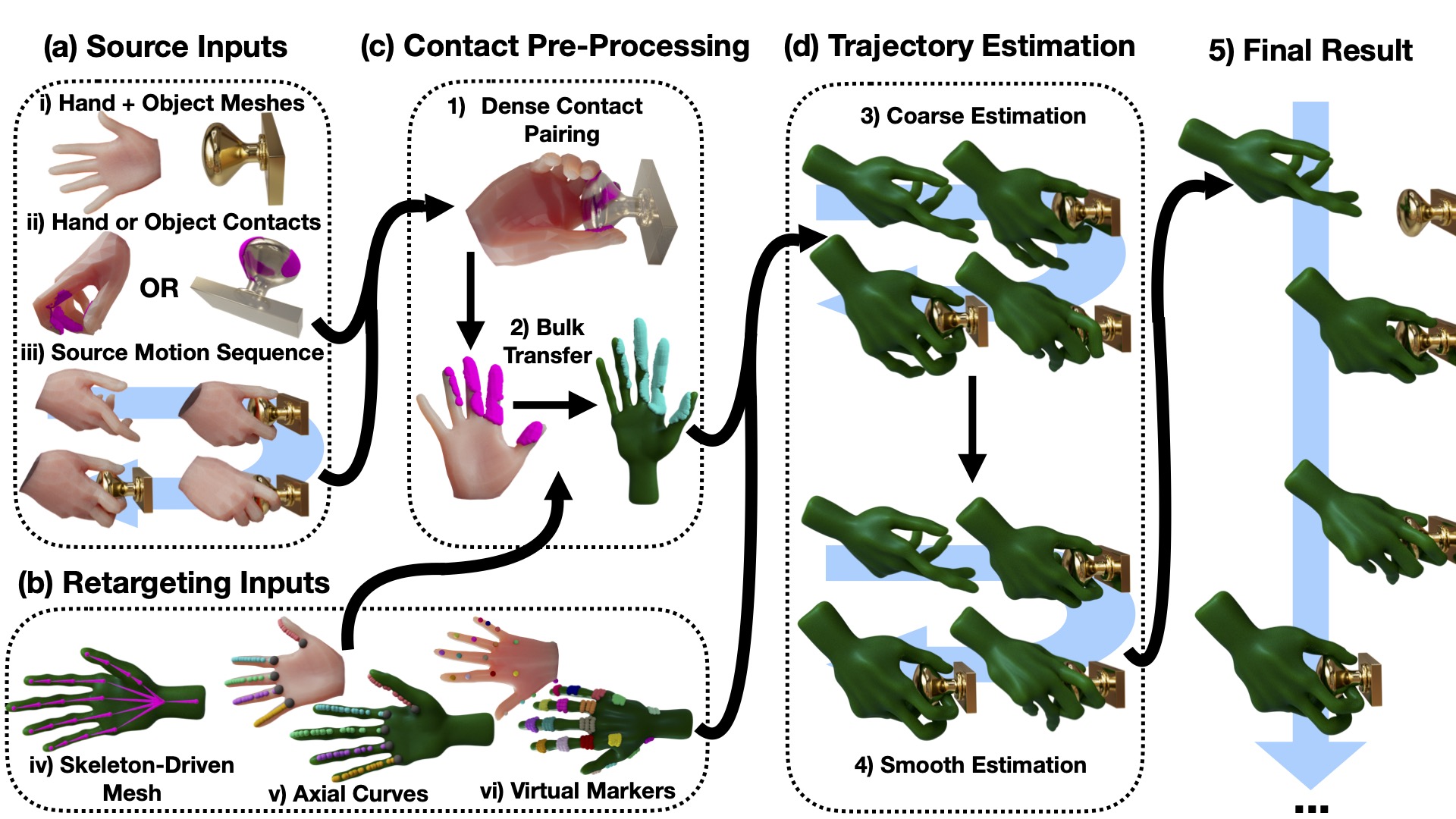}
\caption{Overview of our retargeting framework. (a) Our approach requires inputs of accurate meshes of the original object and source hand, per-frame contacts on either the object or source hand, and a complete motion sequence of the source hand. (b) To perform the retarget, we require a skeleton-driven target hand mesh as well as a set of artist-annotated corresponding virtual markers and axial curves. (c) After recovering a dense set of contacts between the object and source hand, we transfer contacts across the entire time series and (d) use the virtual markers and transferred contacts to synthesize motion for the target hand from scratch.}
\label{fig:overview}
\end{figure*}

Our retargeting pipeline is shown in Figure~\ref{fig:overview} and can be divided into the following steps:

\begin{enumerate}
    \item Extract a dense corresponding set of contact areas between the object and source hand per frame
    \item Procedurally transfer all contacts from the source to target hand across all frames
    \item Estimate an initial trajectory for the target hand using fixed markers and the transferred contacts
    \item Refine the estimates to improve temporal consistency
    \item Construct the final trajectory through spline fitting
\end{enumerate}

\noindent Importantly, our pipeline assumes that the desired solution is one that \textit{attempts to match the interaction mechanics of the source manipulation as exactly as possible}. For this reason, we assume contacts on the object are the same across all hands. We detail the expected inputs and processing steps in the following subsections.

\subsection{Inputs}

Our method requires existing hand-object motion data as input, which we expect to minimally include:

\begin{enumerate}[i]
    \item Accurate meshes of the original object and source hand
    \item A set of dense per-frame contact annotations on either the object or hand mesh
    \item A complete set of frames defining the motion sequence
\end{enumerate}

The GRAB \cite{taheri2020grab} and ARTIC \cite{fan2023arctic} datasets contain all three types of data. We select the former as the data source for our experiments. Importantly, we only require one set of dense contacts and do \textit{not} require the original hand skeleton.

For retargeting to a new hand, we require:

\begin{enumerate}[i]
    \setcounter{enumi}{3}
    \item A skeleton-driven target hand mesh
    \item A set of artist-annotated \textit{corresponding virtual markers} on the source and target hand
    \item A set of artist-annotated \textit{corresponding axial curves} \cite{lakshmipathy2023contactedit} on the source and target hand
\end{enumerate}

\noindent We expand on the required artist annotations and their usages in the following sections. We also require the source and target hands to be manifold; however, it is possible to overcome this constraint by using an approximate manifold wrapper. We demonstrate a result with the fully articulated Allegro Hand$^1$\footnote{$^1$ https://www.wonikrobotics.com/research-robot-hand} as an example. Importantly, we do not assume any sort of morphological or geometric similarity between the source and target hand (e.g. identical finger counts, triangulations, finger or palm shape, finger length).

\subsection{Dense Contact Pairing}
\label{sec:denseContactPairing}

Consistent with existing datasets, we assume contact areas are represented by a finite set of discrete points per frame across the entire motion sequence. We assume a 1:1 correspondence between each hand-object contact point pair. Contact points are stored as barycentric coordinates to render our approach sampling agnostic, and thus robust, to variances in triangulations across different meshes. This approach ensures that data can be collected and transferred from a coarsely triangulated hand (e.g. the MANO hand \cite{romero2017embodied}) to arbitrarily fine or irregularly sampled target hand meshes without risk of discretization error from vertex clamping.

In the event of only one dense set being available, as in the case of the GRAB dataset, we can generate a corresponding dense set through raycasting. We trace contacts out from source mesh locations along element normals until the opposite mesh is intersected, automatically generating a paired barycentric point. In the event of penetration, we invert the trace direction and retry. Points which do not intersect the opposite mesh or are further apart than an $\epsilon$-metric are considered errors and discarded. We note that while this technique proved reasonably reliable in practice, results tend to be inaccurate if the origin manifold exhibits high local curvature at the origin of the traces. Figure~\ref{fig:overview} illustrates an example of dense pairing obtained from raycasting.

\subsection{Hand Shape Matching}

We use the deformed state of the source hand and a (possibly empty) set of dense corresponding contacts between the object and source hand in order to compute the target hand pose per frame. As demonstrated in the grasping literature \cite{lakshmipathy2022contacttransfer,lakshmipathy2023contactedit,wei2023generalized}, we can robustly compute target hand poses during contact-rich frames if we know the target hand contact distribution. To create these contact distributions, we transfer contacts from source to target hands over the entire time series using a one-time sparse set of artist annotations consisting of \textit{virtual markers} and \textit{axial curves}. Virtual markers help to define hand pose in situations where contact information is sparse, and axial curves enable scalable and customizable contact transfer through the use of local charts, as described in Section~\ref{sec:contactAlignment}. Importantly, our input annotation requirements are no greater than existing state of the art methods which exploit user-provided landmark points or curves for non-isometric shape matching \cite{takayama2022compatibletriangulations,gehre2018interactivecurve,aigerman2016orbifold,panine2022landmarkadapted} or keypoint tracking \cite{humbertson2015handson,dasari2023pgdm,qin2022from,sivakumar2022telekinesis,wei2023generalized}, yet they adapt well to dense and rapidly changing contacts. Details of annotations and usage procedures are described below, while Figure~\ref{fig:overview} illustrates the shape matching inputs.

\subsubsection{Virtual Marker Alignment}

\begin{figure}
\centering
\includegraphics[width=1.0\linewidth]{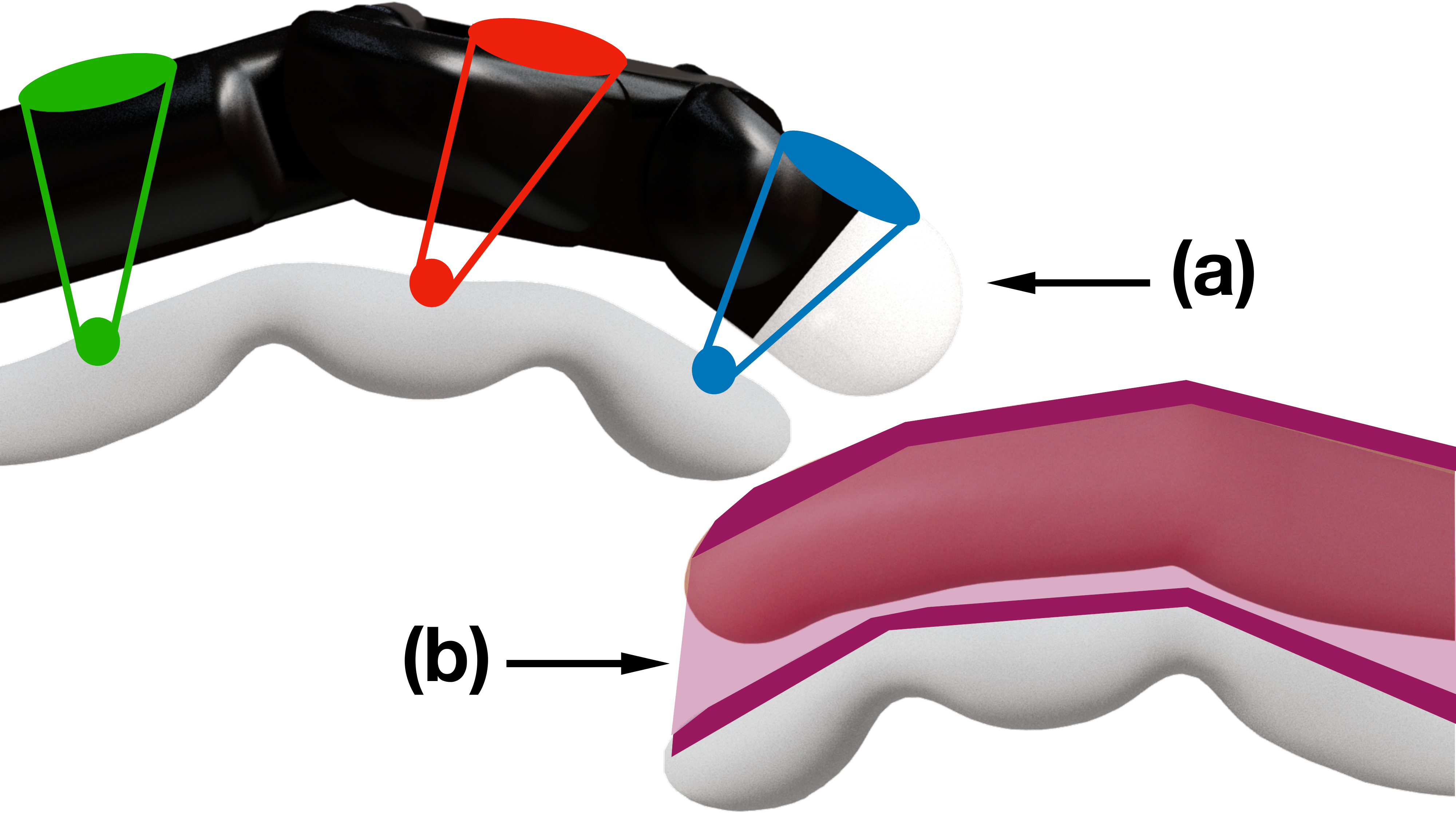}
\caption{Virtual markers can be configured as traditional single-point one-to-one, or alternatively (a) heterogeneous many-to-one, or (b) dense, area-based configurations. Configuration (a) can be utilized to model uncertainty between virtual marker locations between differing source and target hands, which can be useful when finger link lengths are different sizes. Configuration (b) can be used to weight the importance of matching the deformed hand states over large regions, which can be useful when deformation behaviors diverge despite similar link lengths.}
\label{fig:vmalignment}
\end{figure}

Our pipeline expects a corresponding set of artist-annotated virtual markers on both the source and target hand to assist with pose computation, particularly when contact data is sparse or unavailable. We define virtual markers as an arbitrary collection of \textit{fixed} corresponding sets of mesh points between the source and target hand. Markers can either be traditional single points or areas. In the former case, aligning virtual markers for pose computation reduces to traditional keypoint-based IK solving consistent with existing literature \cite{dasari2023pgdm,qin2022from,sivakumar2022telekinesis,wei2023generalized,humbertson2015handson}. Area-based correspondences can be generated by applying existing contact transfer methods \cite{lakshmipathy2022contacttransfer,lakshmipathy2023contactedit} to produce an automatic mapping between such markers, which can be viewed as analogous to matching ``contacts" in mid-air. Notably, however, an area-area configuration requires both sets to contain a 1:1 correspondence of discretized elements. Heterogenous mappings between single point and area based markers are also possible and trivial to designate as a one-to-many association under this modality. Figure~\ref{fig:vmalignment} illustrates each configuration and instances where hetergeneous and area-based correspondences may be beneficial.

\begin{figure}
\centering
\includegraphics[trim={0 7cm 0 8cm},clip,width=1.0\linewidth]{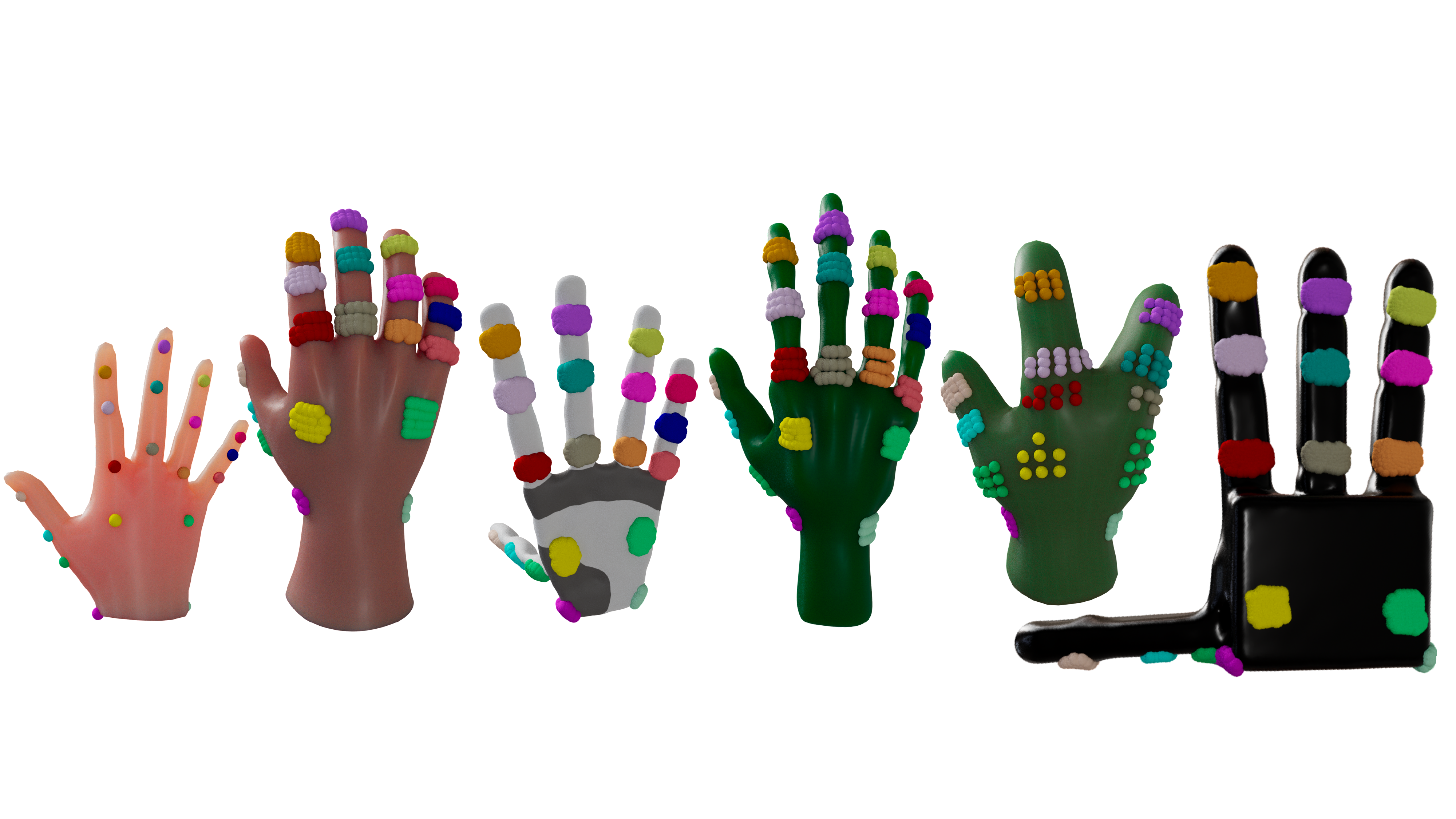}
\caption{Single point virtual marker configuration on the source MANO hand and area based corresponding marker sets on all other hands. We use a manifold wrapper of the Allegro Hand for contact processing.}
\label{fig:virtualmarkers}
\end{figure}

The annotation procedure is manual but was straightforward for an artist and only has to be performed once per source and target hand pair. Details of the annotation process are provided in the appendix. Figure \ref{fig:virtualmarkers} illustrates the heterogeneous virtual marker sets received from the artist and subsequently used as-is for all results, in which the source hand was allocated a set of single-point markers and the target hand an area-based set. In ambiguous cases such as the three-fingered alien hand, the received configuration mapped each human finger marker set to a single finger on the target hand (e.g. thumb, index, middle on the alien hand); however, configurations which map multiple source finger marker sets to a single target finger are supported as well. We also found that area-based virtual markers proved useful in modeling uncertainty on target hands, because it was not always clear where a corresponding single-point virtual marker should be placed.

\subsubsection{Contact Alignment}
\label{sec:contactAlignment}

\begin{figure*}
\centering
\includegraphics[width=0.9\linewidth]{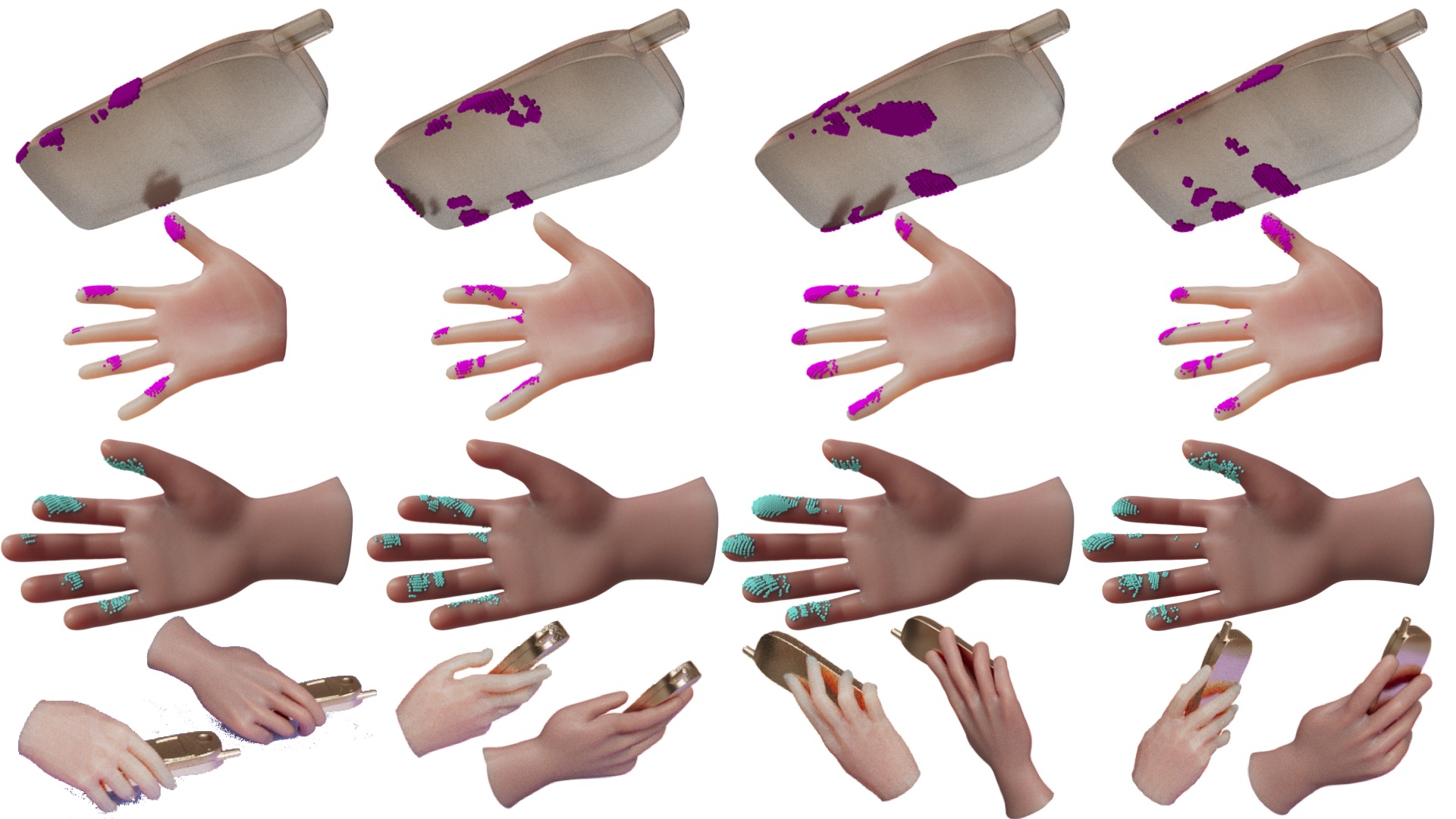}
\caption{(Top Row) Object contacts, (Second Row) source hand contacts, (Third Row) computed target hand contacts, and (Bottom Row) source and retargeted hand motion for four different stages of a complex phone manipulation: (First Column) table pickup, (Second Column) in-hand dialing, (Third Column) holding for use, and (Last Column) movement back towards the table for release. Although poses and contact distributions vary dramatically during the manipulation, our method can successfully produce target hand motion by using source hand contact distributions as a foundational retargeting medium.}
\label{fig:dynamiccontact}
\end{figure*}

Unlike markers, contact distributions are dynamic and can vary greatly between motion frames (Figure~\ref{fig:dynamiccontact}). We model bulk transfer as an intrinsic non-isometric shape matching problem, with the key insight to be formulating the correspondence as being governed by an atlas of multiple coordinate charts \cite{jost2008riemannian}. Atlases can be comprised of one or many coordinate charts, provided the union of all charts generates a cover of the underlying manifold \cite{jost2008riemannian}.

Techniques targeting transfer of global media (e.g. textures), however, typically consider only single chart correspondences and require careful landmark placement to use effectively. Multiple chart correspondences are uncommon in such contexts because careful handling of interpolation across transition regions \cite{jost2008riemannian} and hard chart boundaries is required to maintain global smoothness. Additionally, we found that the inherently high sensitivity to landmark placement makes it challenging for artists to accurately predict and intuitively understand incremental responses to such edits during annotation.

\begin{figure}
\centering
\includegraphics[width=1.0\linewidth]{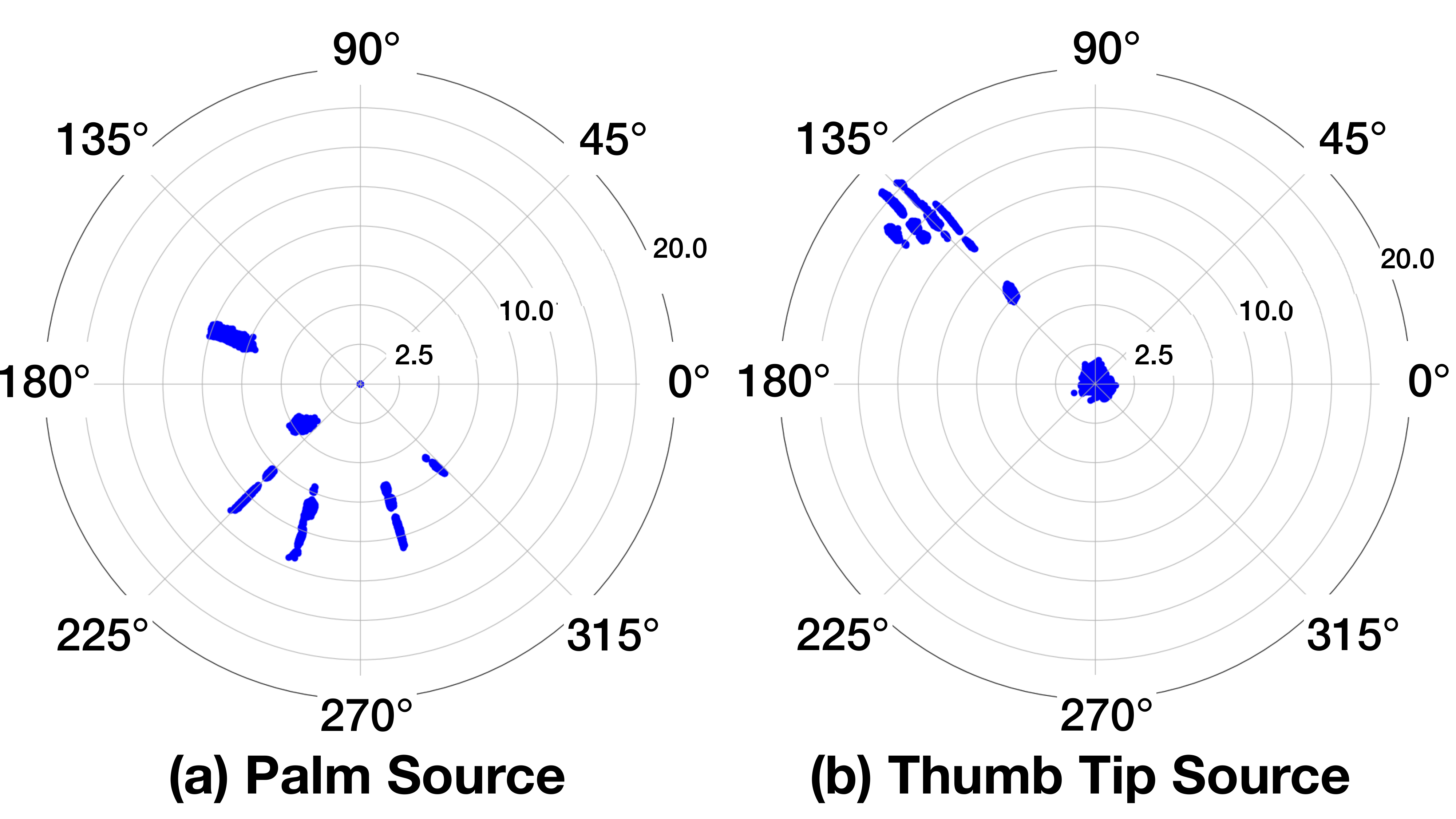}
\caption{(a) When examining geodesic distance and outgoing tangent vector direction from a landmark point on the palm, clusters of contact points are clearly separable. (b) Viewing the same distribution from a landmark on the thumb tip cleanly isolates thumb contacts from those of remaining fingers, allowing us to easily introduce discontinuities to filter out the remaining contacts and subsequently perform arbitrary transformations exclusively on the isolated distribution.}
\label{fig:geodesicgap}
\end{figure}

The highly local nature of contact areas, however, allows us to sidestep many of the drawbacks of mapping global features. For example, a chart used to parameterize contacts on the index finger need not have any influence on a chart used for the middle finger (qualitative semantic boundary), and a chart for contacts towards the tip of a finger need not influence a chart used for contacts towards the base (quantitative geodesic distance boundary). Quantitative boundaries are particularly useful because they can be reliably used even when semantic information is not provided. Figure~\ref{fig:geodesicgap} illustrates such an example, where clusters of contact areas between different fingers are clearly quantitatively distinguishable from each other when viewed from a sample reference point on the palm. It is therefore possible to isolate each finger region by introducing a hard discontinuity in the region of separation, allowing for subsequent fine-grained control over the parameters of each chart independent of the others. The atlas produced from the union of such disjoint charts is thus a segmentation of the manifold; however, unlike typical segmentations of unordered elements, contacts parameterized by any constituent chart can be fully reconstructed through chart inversion. We select the well-understood \textit{logarithmic map} (logmap) as the template for each chart due to its low dimensional parameterization, easy inversion via the \textit{exponential map} (expmap), and ability to be quickly and accurately computed via heat diffusion \cite{sharp2019vhm}. The atlas of source hand manifold $\Omega_S$ is thus formulated as:

\begin{equation}
    \begin{split}
    \mathbb{A}_{\Omega_S} &:= \{(U_{i},\psi_{i}) : i \in [1,M],\ \psi_i = log_{q}(c)\}\\
    s.t.&\ \ \ \bigcup_{i=1}^{M}\ U_{i} = \Omega_S\ \ \ \ \ U_{i} \cap U_{j} = \varnothing\ \forall\ i,j \in [1,M]
    \label{eq:atlassource}
    \end{split}
\end{equation}

\noindent where $(U_{i},\psi_{i})$ represents each of the $M$ constituent charts such that $U_i$ is the subdomain of $\Omega_S$ governed by chart $i$ and $\psi_{i}$ is the logmap function that transforms contact $c$ from barycentric $U_i$ coordinates into logmap coordinates $(r_c,\theta_c)_{q}$ relative to origin $q$ in the transformed space. We then postulate the atlas of target manifold $\Omega_T$ to be of the form:

\begin{equation}
    \begin{split}
    \mathbb{A}_{\Omega_T} &:= \{(V_{i},\psi^{-1}_{i}) : i \in [1,M]\}\\
    &\ \ \ s.t.\ \ \ \bigcup_{i=1}^{M}\ V_{i} = \Omega_T
    \label{eq:atlastarget}
    \end{split}
\end{equation}

\begin{figure}
\includegraphics[width=1.0\linewidth]{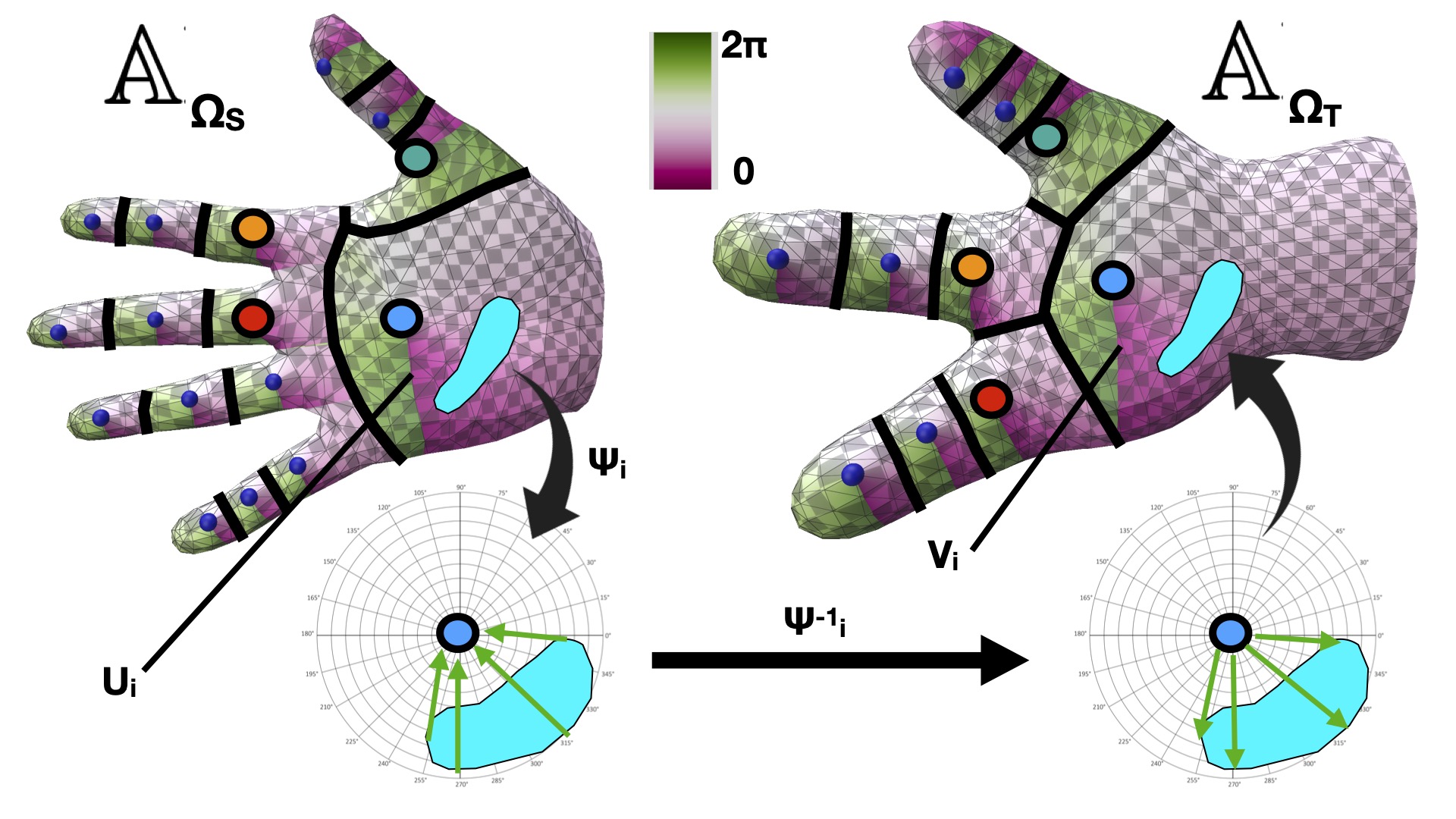}
\caption{Cover generated from a set of landmarks (dark blue), chart boundaries (black), and example corresponding landmarks on each surface (red, orange, teal, light blue) obtained from annotation. A sample contact distribution (cyan) within the boundary of embedded chart $U_i$ is parameterized against the chart's affiliated landmark using logmap transform $\psi_{i}$. Inverting $\psi_{i}$ allows the distribution to be reconstructed from the corresponding landmark on $\mathbb{A}_{\Omega_T}$, while $V_i$ determines the location of the contact's new embedding.}
\label{fig:chartcover}
\end{figure}

\noindent where $\psi^{-1}_{i} = exp_q{c}$, which can be computed by tracing a geodesic originating from $q$ with parameters $(r_c,\theta_c)_{q}$. Including the disjoint condition is not necessary because contact-chart assignment is already determined on $\Omega_S$. Note also that while segmentations of some contact distributions may not always be clear (i.e. full hand power-grasps), any partition of $\Omega_S$ is guaranteed to produce a 1:1 contact point reconstruction on $\Omega_T$. Figure~\ref{fig:chartcover} illustrates the aforementioned terms and proposed formulation.

Our formulation requires only the ability to compute Laplacians and geodesics, which notably are both intrinsic quantities. Unlike extrinsic methods that exploit properties related to a geometry's embedding in space (e.g. vertex locations, normals) \cite{li2007shape}, intrinsic methods instead only consider the connectivity of the structure (e.g. triangle angles, edge lengths). The intrinsic perspective is highly advantageous in the case of hands because such metrics are isometry-invariant, which can allow us to move data between hand geometries at any point in time without knowledge of either the source or target hand pose.

We start by determining each region $U_i \in \Omega_S$ using a set of $M$ landmarks extracted from one-time artist annotation. For convenience, we designate each landmark as the origin ($q$) of each logmap. We next determine the closest landmark to each contact point ($q^*$) using the Vector Heat Method \cite{sharp2019vhm} and MMP \cite{MMP:1987} to extract the logmap coordinates of each contact point $(r_c,\theta_c)_{q^*}$. Importantly, automatically determining $q^*$ means that we \textit{do not require} contact data to be annotated or strictly associated with a pre-determined individual or grouped set of landmarks as mandated in previous work \cite{lakshmipathy2022contacttransfer,lakshmipathy2023contactedit} while also ensuring that each source chart's set of contact points is unique. Notably, while any partition is suitable under our formulation, our approach of taking the closest landmark to each contact under such an atlas is special in that it generates an equivalent point partition to that of a geodesic Voronoi segmentation \cite{herholz2017voronoi} with the added benefit of providing an exact method of reconstruction.

However, rather than requiring landmarks to be provided as individual points, we instead adopt a curve-based input approach \cite{gehre2018interactivecurve}. We select axial curves \cite{lakshmipathy2023contactedit} as the annotation implementation. Axial curves contain:

\begin{enumerate}
    \label{enum:axis-description}
    \item a finite set of points $\{a_1,\dots,a_n\}\in\Omega$ which in our case serve as landmarks, with a \emph{shortest geodesic} $g_i$ connecting each pair of adjacent points $(a_i, a_{i+1})$ for $i=1,\dots,n-1$;
    \item \emph{turning angles} $\{\phi_i\}_{i=2}^{n-1}$, where each $\phi_i$ is the angle of rotation from the ending direction of $g_{i-1}$ to the initial direction of $g_i$, expressed in the tangent space of $a_i$.
\end{enumerate}

\begin{figure*}
\centering
\includegraphics[width=0.9\linewidth]{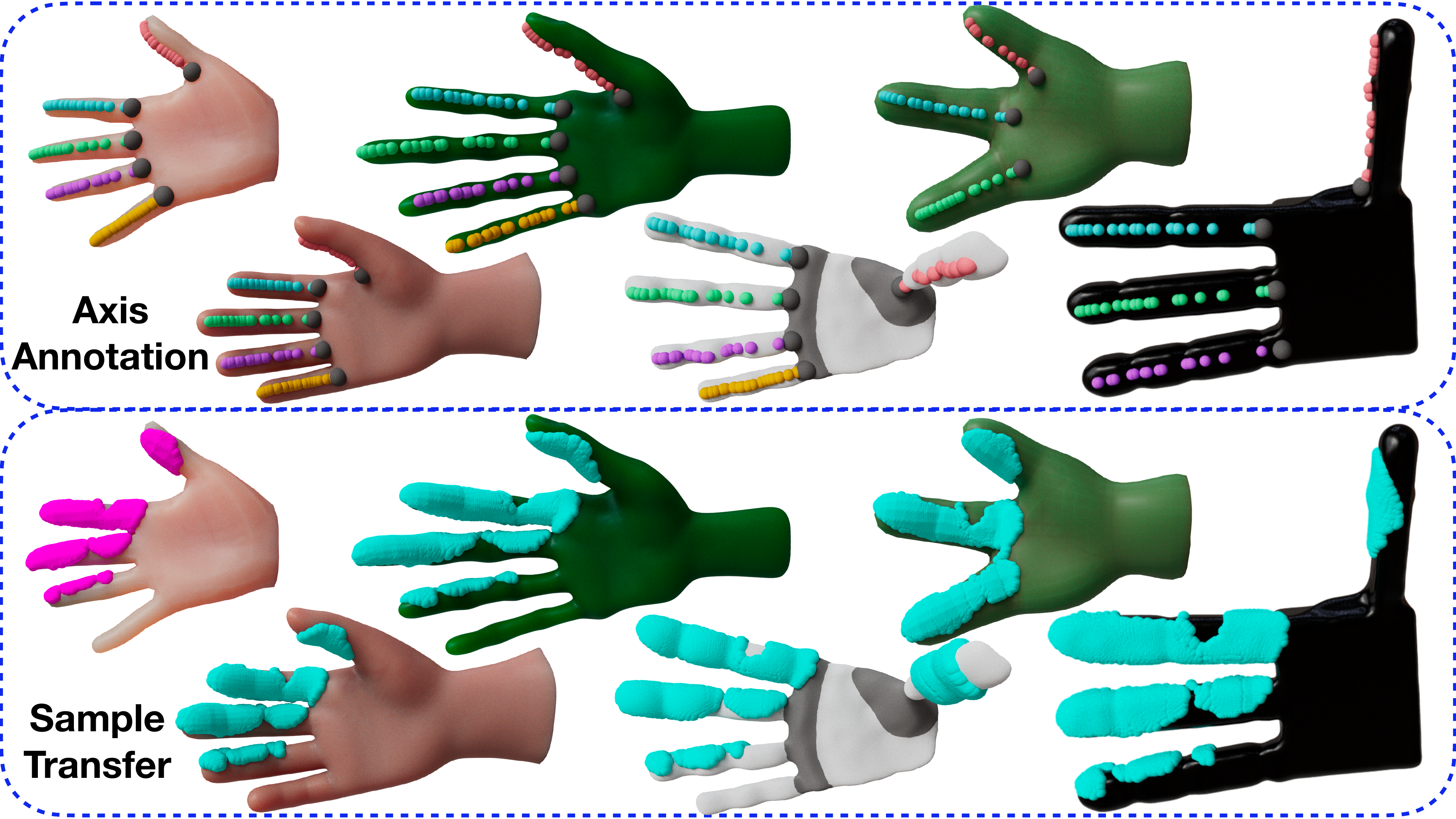}
\caption{One-time axis annotation (top) enables scalable retargeting of original contacts (magenta) to customized configurations per target hand (cyan). Axis colors indicate matching annotations.}
\label{fig:contactretarget}
\end{figure*}

\noindent We found that axial curves proved highly advantageous in enabling our artist to easily generate and tweak multiple landmarks at once via simplified control handles (sparse inputs), thereby greatly reducing annotation time. Axial curves also substantially reduced the need for careful landmark placement, automatically resolved the annotation overhead of consistently orienting the logmap zero angle of all constituent axis points, simplified the process of designating corresponding consistently oriented landmarks on $\Omega_T$, and provided a representation that enabled straightforward modification of expmap reconstruction parameters (discussed shortly). In practice, we found that even simple collections of curves (Figure~\ref{fig:contactretarget}) easily obtainable from an artist yielded remarkably effective atlas correspondences that were capable of transferring contact areas from wildly differing manipulations.

In most typical cases, corresponding axial curves can be easily drawn by an artist; however, there are several interesting alternate cases we can discuss. In the case of $\Omega_T$ having fewer fingers than $\Omega_S$, for example, it is possible to safely ``discard" unwanted contact groups by setting any $V_i = \varnothing$. Similarly, hands with additional fingers do not necessarily require all fingers to be used --- the chart nearest the unused finger simply extends to the region, ensuring a full cover of $\Omega_T$ is still maintained.

\begin{figure}
\centering
\includegraphics[width=1.0\linewidth]{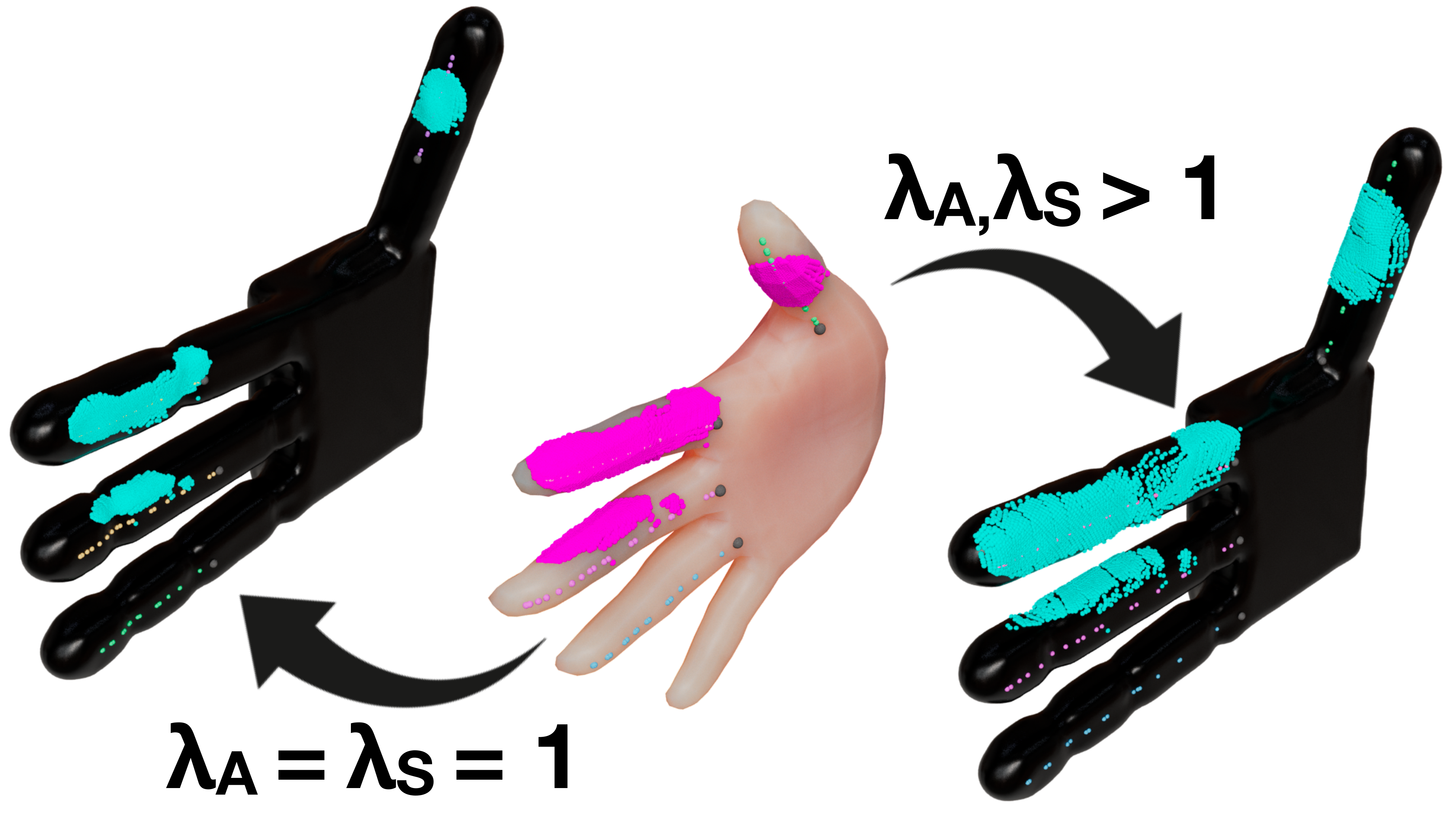}
\caption{Illustration of isometry-preserving-as-possible contact transfer between two widely varying hand shapes. Unnatural squishing of contacts on the target domain fails to model semantic equivalence between fingers.}
\label{fig:badtransfer}
\end{figure}

To address variations in finger or palm shape or girth, we introduce an expmap scaling metric $\lambda_s(\theta)$ and reformulate $\psi^{-1}_i$ as instead tracing geodesics of the form $(\lambda_s(\theta)r,\theta)$. This modification effectively permits ``deforming" contact distributions parameterized on $\Omega_S$ into \textit{any} shape on $\Omega_T$, although in practice we found that uniform scaling was largely sufficient. Although such a parameter breaks the assurance of the expmap providing an isometry-preserving-as-possible reconstruction of contacts embedded by $(U_i,\psi_i)$ on $\Omega_T$, as illustrated in Figure~\ref{fig:badtransfer}, we note that such distortion is often desirable to meaningfully capture \textit{semantic} similarity between geometric variations.

Finally, we address finger length variations by altering geodesic lengths of the axial curve representation to shift relative landmark distances, and by extension the desired locations of $V_i$. We therefore introduce one more parameter $\lambda_a$ that can be used to extend or contract the length of the geodesics connecting the axial curve points on $\Omega_T$. As also illustrated in Figure~\ref{fig:badtransfer}, the combination of $\lambda_s$ and $\lambda_a$ enables fine-grained, predictable adaptation of contact distributions across even widely varying geometries. Importantly, the flexibility to perform such alterations is made possible by the atlas model.

We obtain a corresponding set of landmarks on both $\Omega_S$ and $\Omega_T$, as well as hyperparameters $\lambda_a$ and $\lambda_s$ from one-time \textit{non-expert} artist annotation. We implement and build on tools from existing published work \cite{lakshmipathy2023contactedit} to facilitate the process and detail the annotation procedure in the appendix. Contacts across entire manipulation time series are subsequently procedurally transferred between the source and target hand. Figure \ref{fig:contactretarget} illustrates the received annotations and a representative sample transfer.

\subsection{Motion Retargeting}

We now have all information necessary to synthesize the retargeted motion. As first discussed in Section~\ref{sec:methods}, we perform the reconstruction in three steps: (1) estimation of an initial trajectory for the target hand, (2) refinement of the initial estimate, and (3) construction of the final trajectory through spline fitting. We use a simple and common objective formulation in all steps for computing solutions per frame:

\begin{equation}
    \begin{aligned}
        \displaystyle \vect{\theta_f}^* = \argmin_{\vect{\theta}} \quad & \lambda_{m} \Gamma_{M} + \lambda_{c} \Gamma_{C} + + \lambda_{t} \Gamma_{T} + \lambda_{j} \Gamma_{J}\\
        \textrm{s.t.} \quad & \vect{\theta_L} \leq \vect{\theta} \leq \vect{\theta_U}
    \end{aligned}
    \label{eq:optframe}
\end{equation}

\noindent where $\Gamma_{M}$, $\Gamma_{C}$, $\Gamma_{J}$, and $\Gamma_{T}$ are the penalty terms, $\vect{\theta}$ is the DOF vector, $\vect{\theta_L}$ and $\vect{\theta_U}$ define the lower and upper bounds of $\vect{\theta}$ respectively, and $\lambda_{c}$, $\lambda_{m}$, $\lambda_{j}$, and $\lambda_{t}$ are weighting hyperparameters.

We next describe each penalty term. We start with the marker penalty ($\Gamma_{M}$), which encourages aligning source hand and target hand virtual markers at frame $f$. Assuming $M$ total virtual marker points, the marker penalty is defined as:

\begin{equation}
    \begin{array}{rrclcl}
        \displaystyle \Gamma_{M} &=& \multicolumn{3}{l}{\sum_{m=0}^{M}\ \ \Gamma_{MD,m} }
    \end{array}
    \label{eq:markererror}
\end{equation}

\noindent where $\Gamma_{MD,m}$ represents the $L_2$ distances between corresponding marker points. Next is the contact penalty ($\Gamma_{C}$), which encourages aligning target hand and object contacts at frame $f$. Assuming $C$ total contact points, the contact penalty is defined as:

\begin{equation}
    \begin{array}{rrclcl}
        \displaystyle \Gamma_{C} &=& \multicolumn{3}{l}{\sum_{c=0}^{C}\ \  (\lambda_{cd} \Gamma_{CD,c} + \lambda_{cn} \Gamma_{CN,c})}
    \end{array}
    \label{eq:contacterror}
\end{equation} 

\noindent where $\Gamma_{CD,c}$ represents the $L_2$ distances between corresponding contact points, $\Gamma_{CN,c}$ penalizes deviation from surface normal inversion at the contact points, and $\lambda_{cd}$ and $\lambda_{cn}$ are weighting hyperparameters. We also include a table penalty ($\Gamma_{T}$) that discourages hand-table intersection. Assuming S sampling points on the target hand, the table penalty is defined as:

\begin{equation}
    \begin{array}{rrclcl}
        \displaystyle \Gamma_{T} &=& \multicolumn{3}{l}{\sum_{s=0}^{S}\ \ max(0, -\Gamma_{SD,s})}
    \end{array}
    \label{eq:tableerror}
\end{equation}

\noindent where $\Gamma_{SD,s}$ represents the signed distance function (SDF) of the table evaluated at the location of point s. For simplicity, we use the vertex set of the target hand or its affiliated manifold wrapper as S. We also assume a box geometry for the table, which allows us to compute the SDF analytically$^1$\footnote{$^1$ https://iquilezles.org/articles/distfunctions/}. Lastly, we introduce the ``prior" penalty ($\Gamma_{J}$), which serves as a regularizer against either the default rest pose or the previously existing keyed value at frame $f$. Assuming $J$ DOFs, we obtain:

\begin{equation}
    \begin{array}{rrclcl}
        \displaystyle \Gamma_{J} &=& \multicolumn{3}{l}{\sum_{j}^{J} \Gamma_{P,j} }
    \end{array}
    \label{eq:doferror}
\end{equation}

\noindent where $\Gamma_{P,j}$ represents the deviation between DOF $j$ and its existing value. We use the Method of Moving Asymptotes (MMA) \cite{svanberg2002mma}, a local gradient-based solver, implemented in the NLOPT library \cite{johnson2017nlopt}  to compute solutions to Eq.~\ref{eq:optframe}. 

We perform the motion reconstruction in three stages: initial trajectory estimation, smooth refinement of the initial estimated trajectory, and fitting B-Splines to the smoothed trajectory estimates to recover the final motion. Further details of each step are provided in the appendix.

\section{Results}

Following our commitment to standardization, we use a uniform set of optimization hyperparameters across all hands and motions, the same transfer coefficients for each hand across all motions, and an identical number of control points per motion across all hands and DOFs. We note that no parameter required careful tuning, following our commitments to reliability and simplicity. We also did not modify any of the generated DOF spline motion curves in any of our examples. Please see the supplementary video for results. Parameter values for all hands, transfer coefficients, optimization weighting coefficients, and control point counts are tabulated in the appendix. Result computation times ranged from 4 to 22 hours when run on a single Intel Xeon W-1250 3.3 GHz processor without GPU acceleration. Considerable overhead is currently incurred from finite difference approximation of gradients, which can be reduced through analytical computation.

\begin{figure*}
\centering
\includegraphics[width=1.0\linewidth]{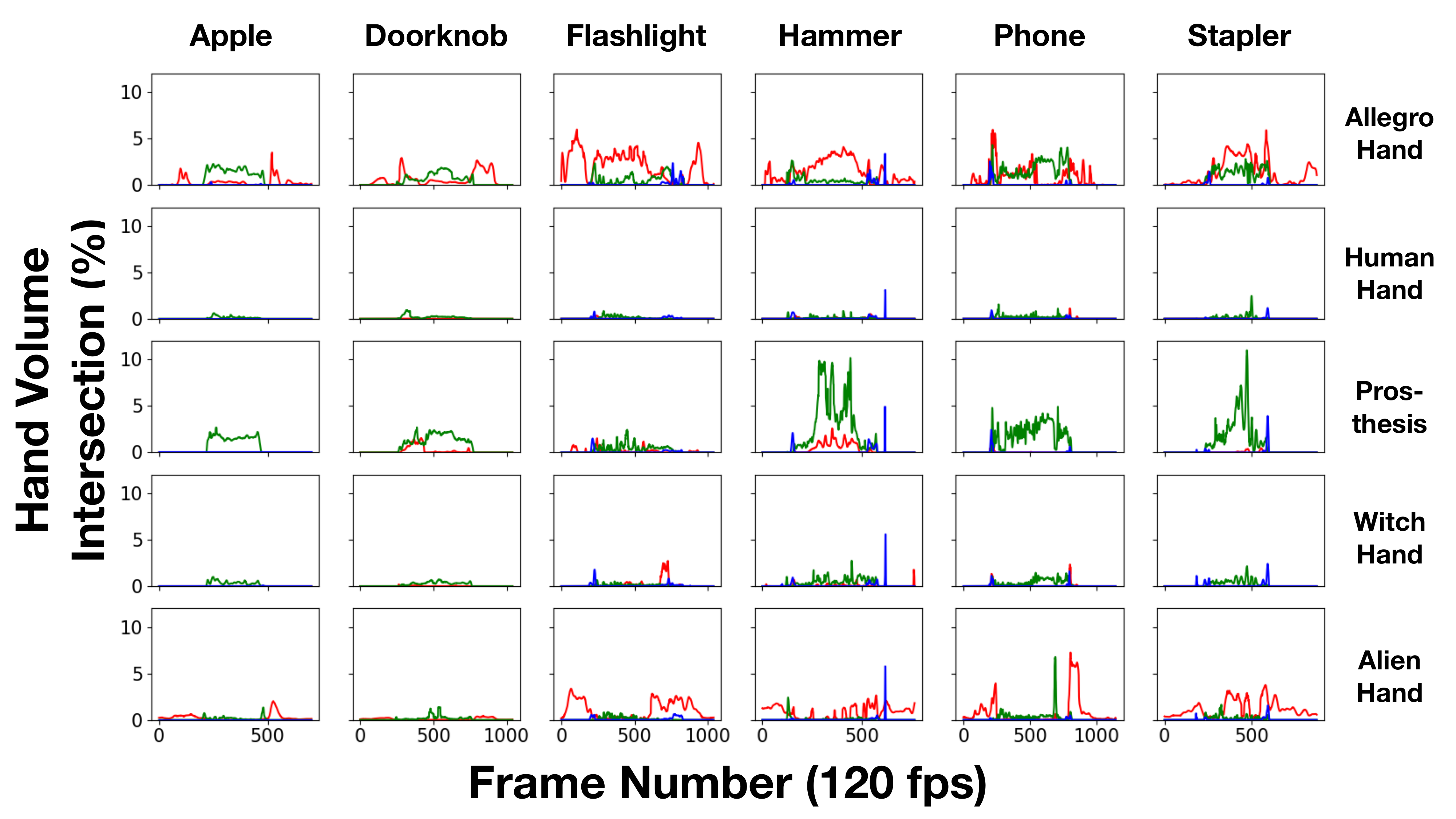}
\caption{Plots of all hand-object (green), self (red), and table (blue) intersection.}
\label{fig:intersectionplots}
\end{figure*}

Figure~\ref{fig:intersectionplots} quantifies the total amount of hand-object, self, and table intersection generated by our method across all results. To compute these quantities, we first determine all penetrating hand vertices via raytracing as detailed in Section~\ref{sec:denseContactPairing}. The resulting points are then clustered via depth-first search on the hand mesh. Finally, we compute the convex hull of each cluster and extract its volume. Total intersection is defined as the sum of all such hull volumes relative to the total volume of the hand, expressed as a percentage. All reported percentages are overestimates due to convex hull approximation. Despite not performing any hand-object or self-intersection resolution, we observe that intersection volumes are nonetheless low. Our results indicate that contact areas, in part due to their implicit encoding of natural grasp states are viable as a cheap approximation of both physical motion and intersection minimization in the absence of a full physics simulation.

Overall, we found the retargeted trajectories to be of generally high quality despite blanket standardization. Such generalization is particularly notable since results confirm that parameterization of contacts across arbitrary motions is possible \textit{without knowledge of distributions in advance} and without customization of landmarks to individual distributions.

\section{Extensions}

We next discuss two extensions of our method beyond hand motion retargeting: hand design visualization and retargeting demonstrations to different objects.

\subsection{Visualizing Design Choice Impacts}

\begin{figure}
\centering
\includegraphics[width=1.0\linewidth]{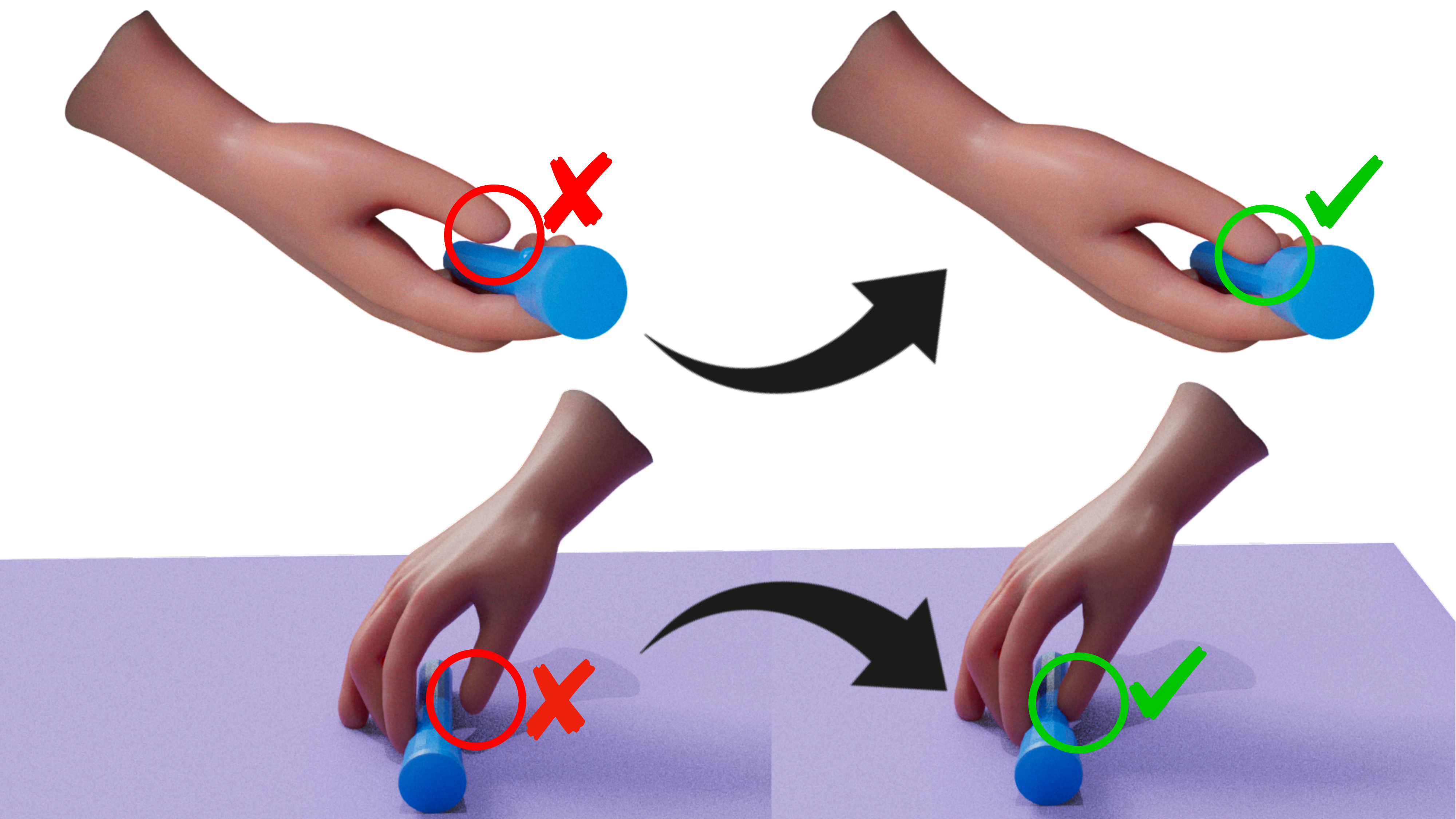}
\caption{(a) The original design candidate thumb is unable to reach the flashlight switch while (b) the revised candidate thumb length adequately closes the gap in all such frames}
\label{fig:designchange}
\end{figure}

\begin{figure}
\centering
\includegraphics[width=1.0\linewidth]{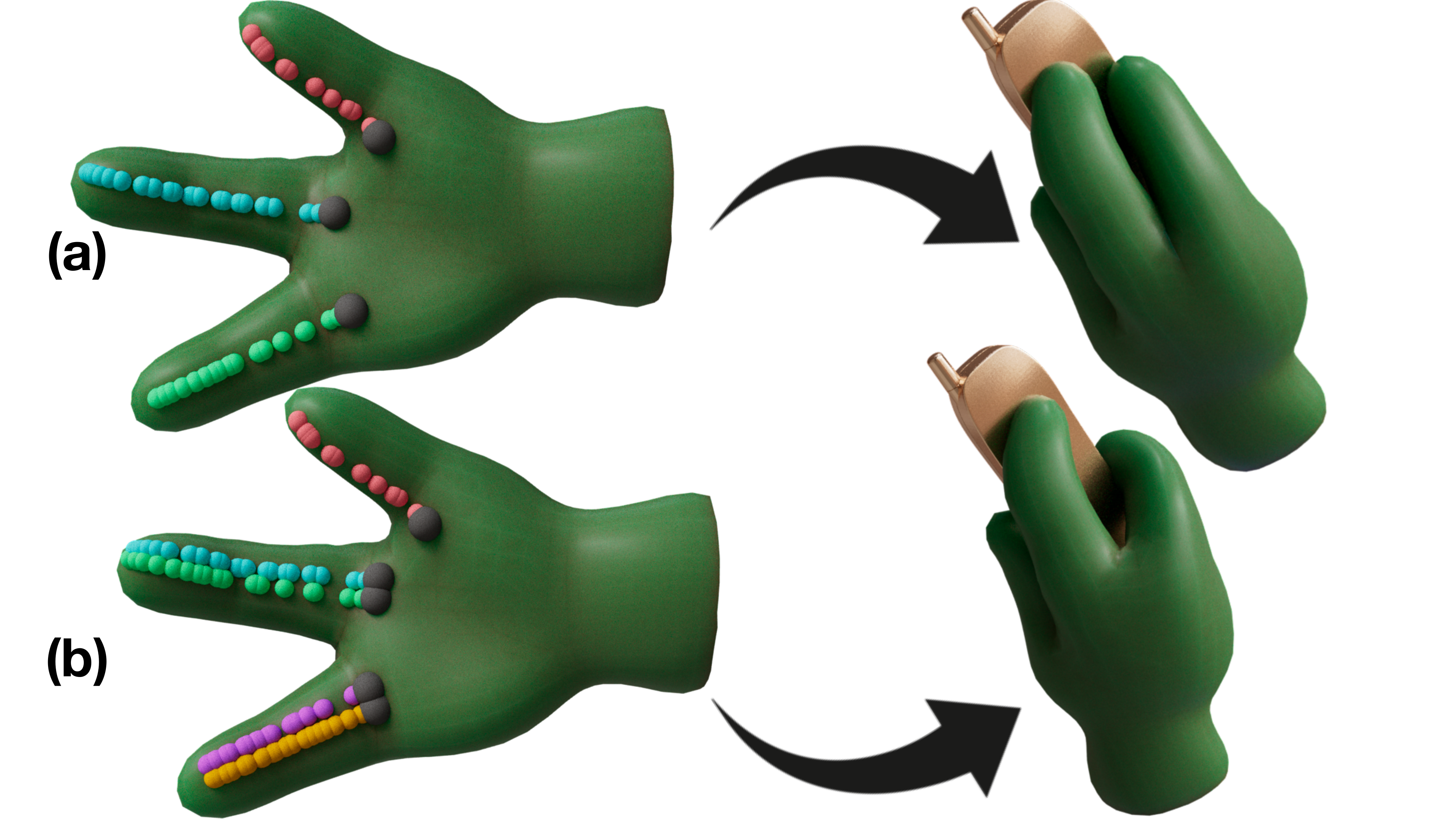}
\caption{(a) A typical three-finger axis mapping and its resulting pose compared to (b) an unusual ``Spock"-like configuration achieved by assigning multiple curves to a single finger.}
\label{fig:yodamultiple}
\end{figure}

Our method enables task-specific visualization of design choices as well as insights into how such parameters can be adjusted. For example, determining appropriate finger lengths is a common problem in both rigging and rapid prototyping contexts due to the decoupling of design and usage. Figure~\ref{fig:designchange} illustrates an example in which the thumb in the original design is not able to reach the object when the source contact distribution is used, while the revised candidate is able to do so. Full trajectories can be viewed in the supplementary video.

Another such example is testing alternate finger mappings. As illustrated in Figure~\ref{fig:yodamultiple}, it is possible to change the three-fingered hand's interaction with the phone by creating multiple axial curves on a single finger, effectively semantically mapping contacts from two human fingers (e.g. index and middle) to a single alien finger. Such flexibility is particularly useful for hands with different finger counts or other morphological differences.

We also provide a third example demonstrating the ability of our approach to handle changing the number of DOFs in the appendix.

\subsection{Object Substitution}

\begin{figure}
\centering
\includegraphics[width=1.0\linewidth]{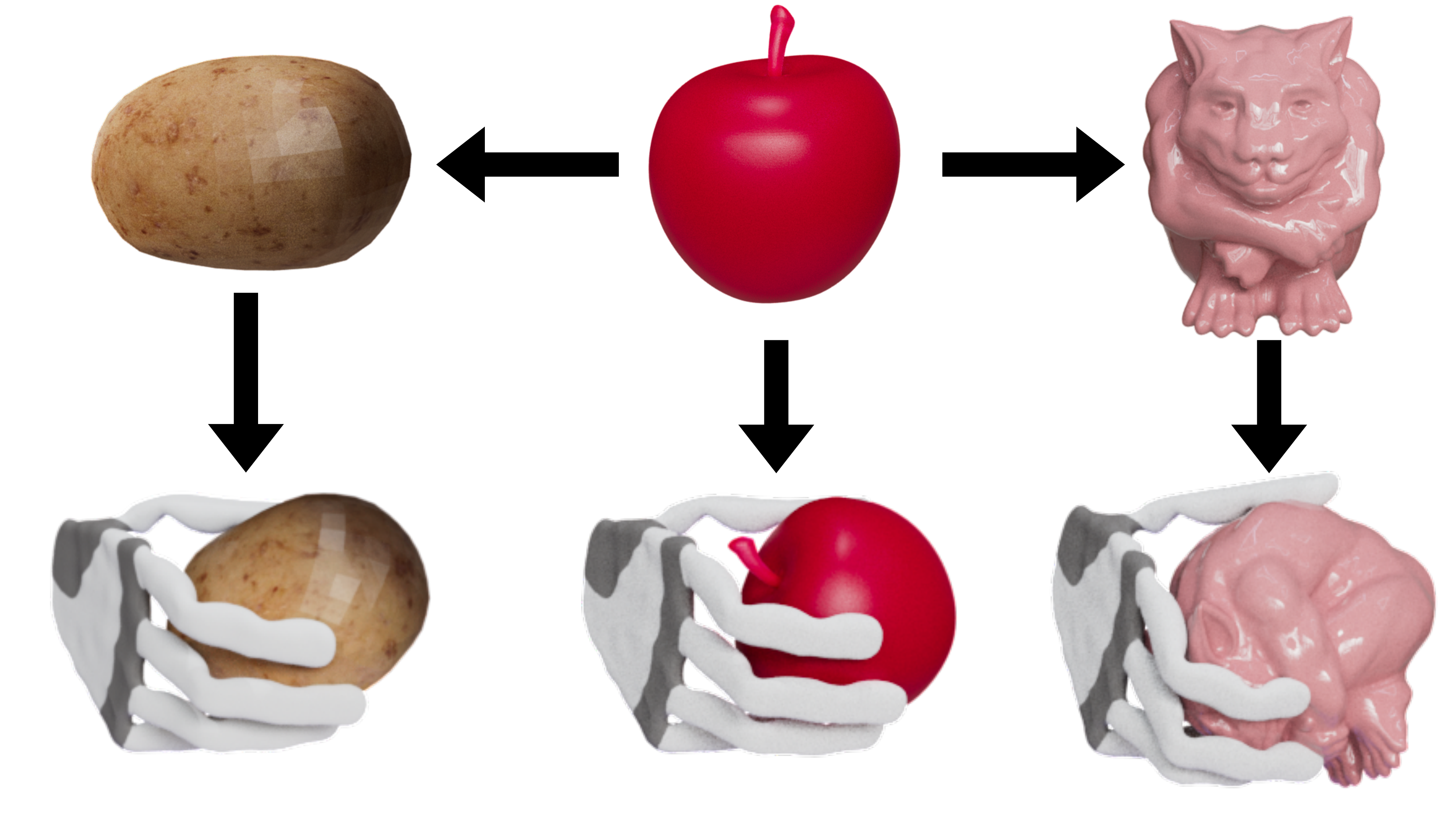}
\caption{Our method can be extended to  accommodate retargeting generic motions between different objects. Hand grasps successfully make subtle, but important adjustments to adapt to simple shapes (left) as well as more dramatic adjustments for more complex features (right).}
\label{fig:objecttransfer}
\end{figure}

\begin{figure}
\centering
\includegraphics[width=1.0\linewidth]{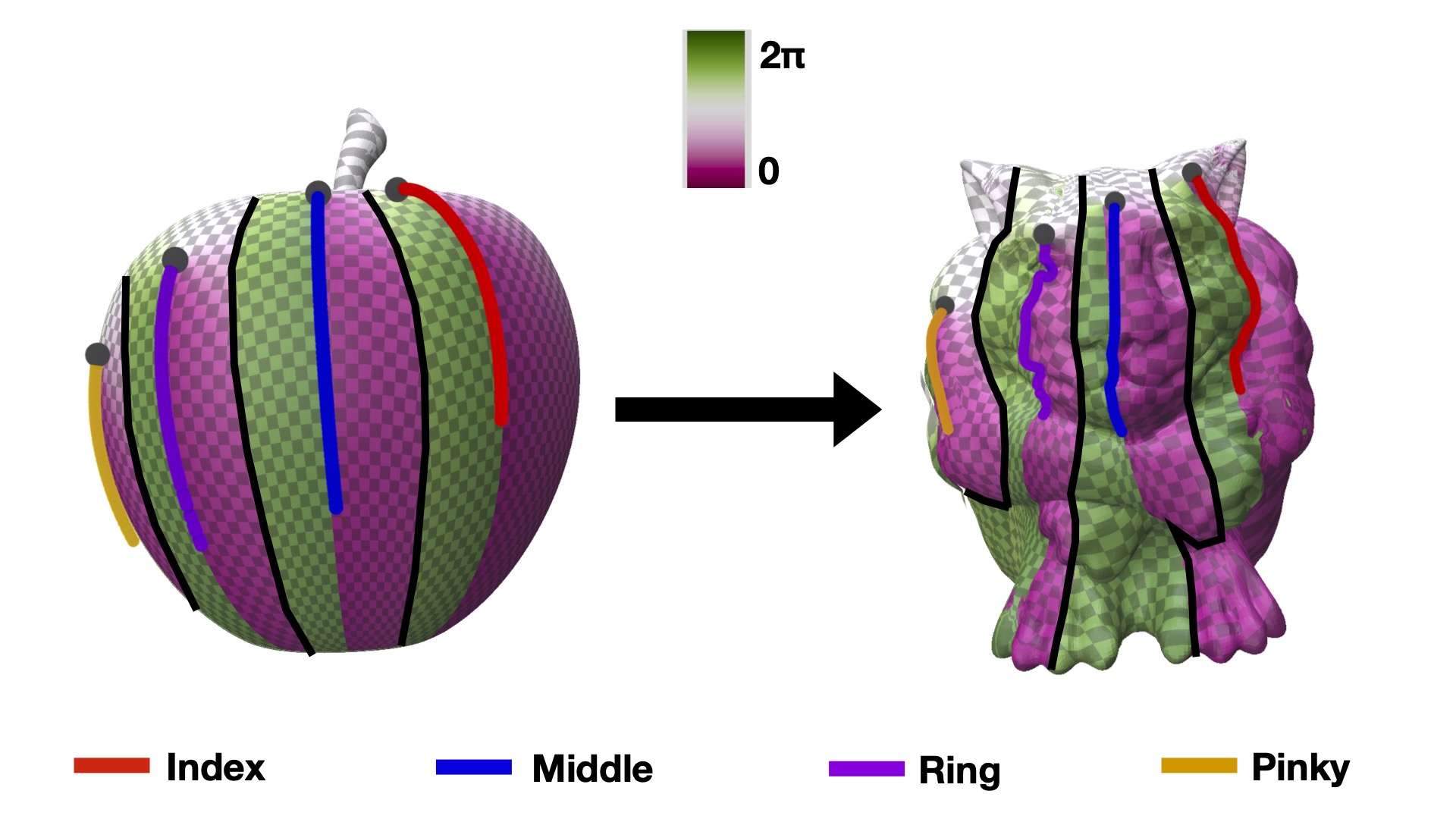}
\caption{Approximate atlases used to retarget contacts of a handoff manipulation between a simple apple and a more geometrically complex gargoyle figurine. Axial curves corresponding to each finger are denoted by different curve colors, while black spheres denote the start point of each curve. Curve placements on the target surface can be arbitrary and be used to implicitly modify grips on the target surface to any extent desired.}
\label{fig:objectatlas}
\end{figure}

Our method also enables retargeting generic manipulations to different objects. Contact sets can be generated on new objects through any means of correspondance, including by reversing the raytrace process from the source hand, tracing from the original object, through diffusion models \cite{wei2023generalized}, or, perhaps most interestingly, by using the same atlas-based approach used for blanket retargeting of hand contacts. The remainder of the pipeline is subsequently applied normally to produce the final result. We illustrate an example in both Figure~\ref{fig:objecttransfer} and Figure~\ref{fig:objectatlas} and in the supplementary video of re-purposing an existing apple ``handoff" manipulation to a potato, where contacts on the potato are generated by tracing the original contacts outward from the apple, and a more geometrically complex gargoyle sourced from the Thingi10K dataset \cite{zhou2016thingi10k}, where contacts are instead retargeted to slightly different configurations using an atlas between the apple and the gargoyle.

\section{Comparisons}

We next validate the importance of contact information by comparing our approach against two existing contact-free methods used in tele-operation: fingertip keypoint tracking \cite{qin2022from,dasari2023pgdm} and functional pose equivalence \cite{Handa2019DexPilotVT}. We use the Allegro Hand for consistency with the baselines. To isolate the objective formulation, we beneficially augment both methods with information about the known full source trajectory. Specifically, we provide access to ground truth root joint estimation, independent pose computation per frame (Section~\ref{sec:initialtrajectoryestimation}), and retroactive smoothing of the whole sequence (Sections~\ref{sec:trajectoryrefinement}-\ref{sec:splineFitting}).

\subsection{Fingertip Keypoint Tracking}

We first consider the popular strategy of fingertip keypoint tracking. We select four keypoints at the tips of each Allegro finger and assign them to track the human thumb, index, middle, and ring fingertip positions. The corresponding Allegro hand pose is computed using an existing optimization formulation \cite{qin2022from}, which is equivalent to Eq.~\ref{eq:optframe} and \ref{eq:optsequence} with $\lambda_C = 0$.

\subsection{Functional Pose Equivalence}

We next consider the strategy of generating an equivalent functional pose on the target hand, which instead considers relative distances between pairs of keypoints on each hand and measures equivalence in terms of task space keyvectors \cite{Handa2019DexPilotVT, sivakumar2022telekinesis}. Specifically, we use the objective formulation proposed by DexPilot \cite{Handa2019DexPilotVT} as the basis for pose computation. Exact keyvectors and detailed term explanations are available in the cited work.

\smallskip

\begin{figure}
\centering
\includegraphics[width=1.0\linewidth]{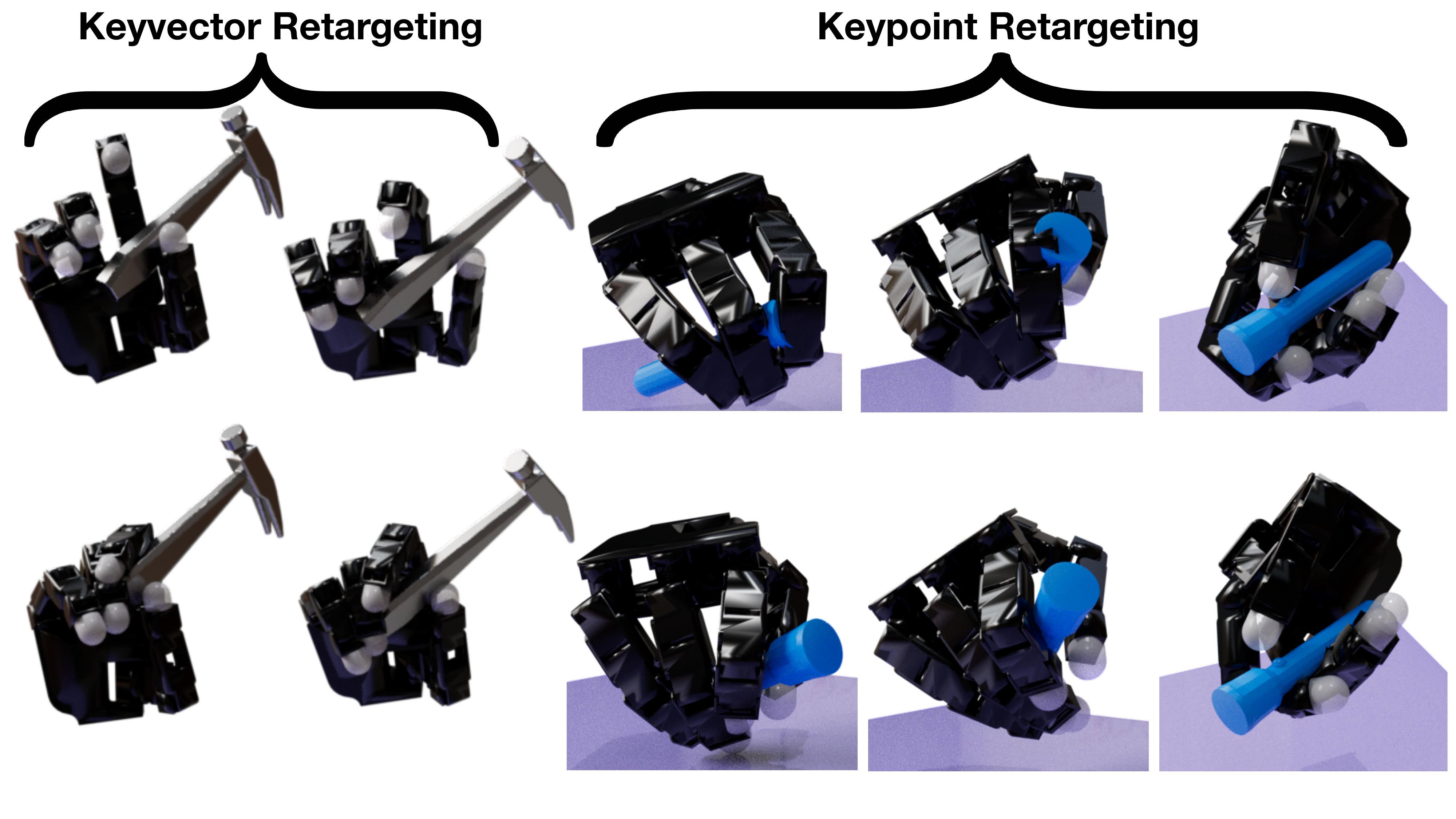}
\caption{(Top row) Functional pose equivalence in terms of keypoints or keyvectors visibly struggle encouraging hand-object contact and motion alignment especially when source and target hand geometries greatly diverge. (Bottom row) Using contact areas greatly reduces such artifacts.}
\label{fig:methodcomparison}
\end{figure}

Full trajectory comparisons to both methods are available in the supplementary video. As illustrated Figure~\ref{fig:methodcomparison}, we observe that both methods, while capable of producing smooth trajectory estimates, generally struggle to make contact with the object and produce non-trivial motion misalignment artifacts. Because both of these techniques have been used for retargeting in teleoperation scenarios, a person in the loop can interactively correct for errors such as those shown in Figure~\ref{fig:methodcomparison}.  In contrast, the approach presented in this paper provides results with good contact without the need for such intervention.

\section{Discussion}

\begin{figure}
\centering
\includegraphics[width=1.0\linewidth]{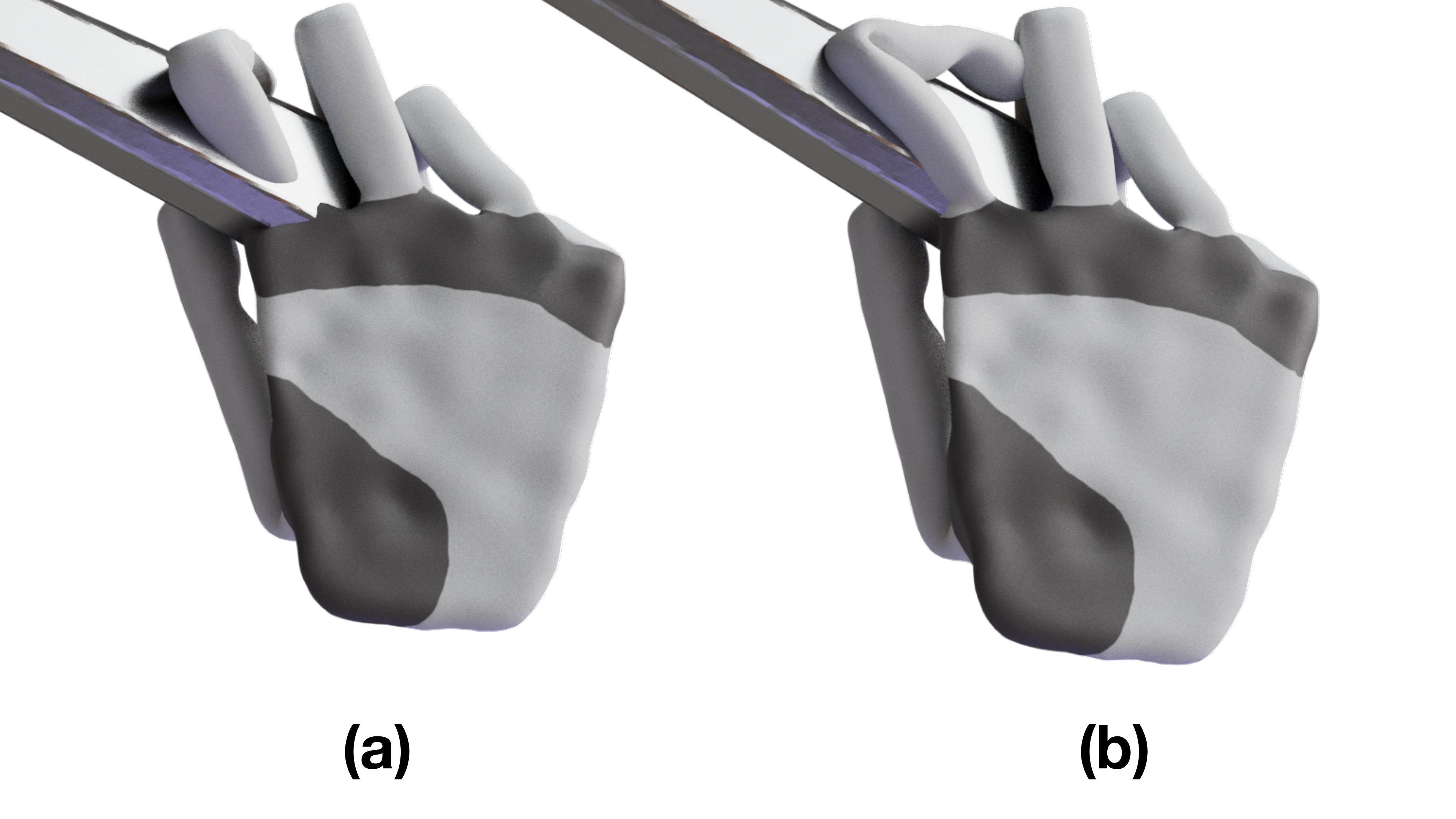}
\caption{When optimized from the same start position with the same marker and contact distribution targets, (a) a 2-DOF MCP joint converges to a solution which penetrates through the hammer while (b) a 3-DOF MCP joint is able to conform to the hammer's handle with much less intersection. We found that the torsional DOF was utilized in this particular grip across all hands that included a 3-DOF MCP joint.}
\label{fig:prosthesishammer}
\end{figure}

We were pleasantly surprised to find that high quality retargets were possible even with a uniform set of parameters across all hands, although certain hands were comparatively easier to retarget than others. Surprisingly, the three fingered alien hand was the easiest to work with despite its unusual morphology. Its comparatively thick size was highly beneficial for limiting pose estimate uncertainty with respect to both markers and contacts. The only noticeable drawback was higher relative self-intersection compared to other hands; however, this behavior was unsurprising since the large majority of these intersections were caused by overlap between the second and third phalange, which were mapped to to the index and middle finger of the human hand --- two fingers which commonly moved together in close proximity within the task suite we examined. In contrast, despite its more anthropomorphic appearance, estimations of the prosthesis wrist position were comparatively far noisier due to its thin profile. The prosthesis also exhibited relatively high hand-object intersection for the hammer and stapler tasks, which we found could be attributed to its small size and limited DOFs. In the case of the stapler, the prosthesis had difficulty curling fingers cleanly around the entire object. In the case of the hammer, as illustrated in Figure~\ref{fig:prosthesishammer}, the lack of a torsional MCP DOF caused the index finger to penetrate through the hammer rather than wrapping around the handle. The witch and Allegro hand were the most difficult to retarget due to their unusually long finger lengths, which we argue is one of the most significant factors in determining retargeting complexity between hands.

Additionally, although we focus on hands in this paper, we showed that our shape matching approach is general and can be used to precisely define transfer of arbitrary local data between shapes. Additional applications include defining complex interactions between surfaces (e.g. between two deformables), adaptable manipulation planning, or movement of entire assets positioned on one surface to another for surface animation retargeting. We are interested in exploring additional extensions of our approach to broader contexts in the future.

Finally, we note that we did not receive any hand pairs following an area-area virtual marker configuration; however, because our goal in this work was to examine a wide variety of dissimilar hands, we suspect that the annotator's decision to use only many-to-one configurations is largely a result of experiment specifics. Even the retargeted human hand we used had different finger length sizes than the source, and the source hand had an unknown kinematic structure. Retargeting between human hands of similar proportion or highly anthropomorphic robot hands (e.g. the Shadow Hand$^1$\footnote{$^1$ https://www.shadowrobot.com/dexterous-hand-series/}) may provide a more compelling use case for area-area configurations.

\section{Drawbacks and Limitations}

While effective in retargeting motions across a wide range of hands, objects, and manipulations, our method contains several limitations.

The atlas generated by our shape matching approach is notably discontinuous and non-differentiable, making it unsuitable for gradient computations or transfer of global media such as smooth functions or textures. Although these properties were not critical to our problem scope, formulating a $C^{\infty}$ atlas would potentially enable a broader range of useful applications.

Our method does not guarantee a retargeted solution free of artifacts. The most common artifacts were wobbles resulting from leftover trajectory estimation noise or under / over fitting splines. However, both of these artifacts can generally be resolved by tweaking the acceleration cutoff for trajectory refinement or adjusting the number of control points. We found that a fixed set of parameters was suitable for most motions, but note that further refinement is possible with per-hand-per-motion customization if desired.

Our method sometimes has difficulty resolving table intersections in a manner that appears natural, which can result in uncanny finger contortions during object pickup and release. This issue arose in the flashlight and stapler manipulations by the witch hand, and the problem is generally more common in unconstrained long-fingered hands. It is possible to mitigate this issue by constraining or removing DOFs, assuming skeleton modification is permissible.

Finally, our formulation of matching object contacts from the source hand is generally less effective for hands with large morphological or kinematic divergence from the source hand. The Allegro hand is a prime example of a challenging retargeting problem because of its size and limited DOFs. The main difficulty caused by the large size of this manipulator was greater self-intersection, the likelihood of which is generally higher in larger hands. In such instances, and in others where the target hand diverges enough to substantially alter the general expected interaction mechanics (e.g. a human vs. a Barrett Hand$^2$\footnote{$^2$ https://robots.ros.org/barrett-hand/}) or with respect to a particular manipulation (e.g. a pinch grasp on a larger hand vs. a human power grasp), matching original human contacts exactly may not be the best problem formulation. Addressing such divergences would be an interesting area for future work.

\section{Conclusion and Future Work}

We have presented a simple, reliable, and standardized framework capable of kinematically retargeting contact-rich manipulations to a wide variety of target hands. Central to our method is the utilization of contact areas, for which we have presented both a novel, atlas-based shape matching algorithm capable of transferring localized point data procedurally with high control and precision, and an optimization pipeline capable of utilizing said information to create high quality retargeted results. We have shown that our method is capable of enabling unique extensions, including object substitution and visualizing the impact of design parameters. Finally, we have shown the value of contact information and key aspects of our processing pipeline through baseline comparisons.

A natural next step would be to extend our currently purely kinematic retargeting framework to incorporate contact dynamics. However, a crucial limitation is that many existing contact estimates, including those from the dataset serving as the ground truth for this work, are typically determined exclusively from distance-based metrics. We are curious to explore whether collecting contacts in the real world via tactile sensing can be used to simultaneously ground force estimates from inverse dynamics as well as improve the quality of contact localization.

Additionally, contact-driven methods are currently an under-explored topic particularly within the reinforcement and imitation learning literature. We are curious to see if such information can improve policy learning through the reduction of convergence times, encouragement of natural behaviors, or enabling of manipulations not previously possible. Ultimately, we are curious to see if collection and online perception of contact data can enable human to robot transfer of complex manipulations in the real world.

\bibliographystyle{ACM-Reference-Format}
\bibliography{main}

\newpage

\appendix

\section{Artist Annotation Procedures}

\subsection{Virtual Markers}

We provided the artist with a representative photo of a full hand marker set commonly used in motion capture. We requested the artist to designate equivalent sets of markers on the source and all target hands. The artist performed the annotation by selecting individual or groups of vertices on each mesh using default brush tools commonly available in 3D modeling software. The received annotations were processed and stored as corresponding virtual marker sets for all experiments.

\subsection{Axial Curves}

We requested the artist to identify equivalent geometric features on the palmar surface of each hand, but did not provide any directions regarding what said regions included.

We provided the artist a custom tool set to assist with the axis creation process. To do so, the artist first selects two points on the source hand. A geodesic is then traced between the two points in the order of selection, and all intermediate edge crossings are automatically designated as discrete axis points. Traversal between points is stored via the exponential map with the turning angle of the outgoing tangent vector being computed relative to that of the incoming tangent vector. This representation allows for approximate isometric reconstruction independently of the tangent basis at any point except the starting location, which is critical for the transfer procedure. A more comprehensive explanation of the axial curve model is available in existing work \cite{lakshmipathy2023contactedit}.

The artist then selects two points on the target hand to initiate the transfer process. We use the first point as the origin for the axis reconstruction and the geodesic trace between the two points to determine the initial outgoing turning angle. The remainder of the axis curve is then reconstructed automatically from the exponential map computed on the source domain. After reconstruction, the artist proceeds to determine an appropriate $\lambda_A$ for the target hand axis. In practice, a satisfactory $\lambda_A$ extended the axis along the full length of the target finger. The procedure is repeated until all axial curves have been designated on the source hand, transferred to the target hand, and fitted with an appropriate $\lambda_A$. The artist subsequently re-used the annotated source hand to accelerate the annotation of the remaining target hands.

To determine $\lambda_S$, we provided the artist three sample frames of contact data on the source hand from a single motion sequence. Contacts were first loaded and parameterized on the source hand. Subsequent hands then re-used the parameterized representation to accelerate the annotation procedure. After parameterization, contacts were transferred to the target hand and reconstucted from corresponding axial points under the linear scaling previously determined by $\lambda_A$. $\lambda_S$ was then adjusted until an appropriate semantically meaningful distribution was produced. The artist then validated the parameter choice by repeating the transfer procedure with the remaining two frames of data. The procedure was repeated with all axial curves and hands to complete the annotation. The determined parameter values were used to procedurally transfer all frames of contact data in all result manipulations without any additional adjustment.

\section{Motion Retargeting Details}

In the proceeding subsections, ``solving" and ``re-solving" both refer to computing the optimal solution to Eq.~\ref{eq:optframe} using MMA \cite{svanberg2002mma}.
 
\subsection{Initial Trajectory Estimation}
\label{sec:initialtrajectoryestimation}

Since no baseline data is available for the target hand motion, and we cannot easily obtain a preliminary trajectory through direct joint retargeting, we must synthesize it from scratch. We do so by first estimating a per-frame optimal initial trajectory estimate over $F$ total motion frames, which we define as:

\begin{equation}
    \begin{array}{rrclcl}
        \displaystyle \vect{\Theta}^* = \{\vect{\theta_0}^*, \vect{\theta_1}^*,...,\vect{\theta_F}^*\}\\
    \end{array}
    \label{eq:optsequence}
\end{equation}

Importantly, the above formulation entails that each frame of the optimal trajectory \textit{is independent of the estimates of its neighbors,} which runs counter to many existing works that add explicit conditioning terms on the estimate of the previous pose \cite{Handa2019DexPilotVT,qin2021DexMVIL,sivakumar2022telekinesis}. This crucial distinction proved vital in mitigating error buildup over long sequences, which otherwise caused the estimates of frames later in the motion to converge to highly undesirable local minima. Independent estimation also allowed us to reliably prune poor locally optimal estimates during later processing stages.

We perform the trajectory estimation in two passes. First, we solve for only root joint position per frame while keeping the remaining joints fixed in their default positions. We found during our experiments and baseline comparisons that the root joint pose is very important and has a disproportionately large impact on retargeting success. We seed the current frame estimate from the optimal solution of the previous frame exclusively at this stage for convenience. Empirically, we found that solutions produced from such seeding did not vary substantially from those with cold starts. We then solve for the full pose per frame in the second pass using the root estimate as the seed. Default poses for all hands are illustrated in Figure~\ref{fig:allhands}.

\subsection{Trajectory Refinement}
\label{sec:trajectoryrefinement}

We next refine the estimated trajectory to improve temporal consistency using finite acceleration as the smoothing metric. To do so, we impose a threshold $\mathcal{E}_{acc}$ and remove all frames that violate the threshold. We then replace each violated frame with a linear interpolation between its nearest valid left and right neighbors and re-solve. Frames are removed on a per-DOF basis and in a single pass to ensure independence of computation when re-solved. We impose the same threshold across all DOFs in our results regardless of joint coupling (e.g. co-dependence between ball joint DOFs) or units (e.g. angular vs. linear DOFs) in support of standardization; however, different cutoffs can be applied on a per-dof basis if needed.

However, updating violations can introduce new violations in the resulting trajectory. We thus perform the refinement procedure iteratively until either until no further violations are found or all iterations are exhausted. All unresolved violations at the end of iteration exhaustion are ignored during fitting.

Applying this refinement strategy to the raw initial trajectory slowed the method considerably in practice due to high numbers of violations resulting from independent pose computation per frame. Instead, following precedent \cite{Handa2019DexPilotVT,sivakumar2022telekinesis,qin2022from}, we found it beneficial to re-solve for the estimates after applying low-pass and peak removal filters. Pre-processing substantially reduced the number of acceleration violations, and thus the overall computation time required for this stage.

\subsection{Spline Fitting}
\label{sec:splineFitting}

Finally, we use the refined frame solutions as sample points and fit a cubic B-spline to each DOF across the computed time series, ensuring our final solution guarantees $C^2$ continuity. We designate a fixed number of control points and solve simultaneously for both values and locations in time using least squares pseudo-inverse approximation \cite{eberleyBsplineFit}. This representation also permits artist control over the fitted spline, and better approximations, at the cost of possible overfitting, can be easily obtained either by increasing the number of control points or hierarchically compositing splines to reduce error residuals \cite{lee1999hierarchicalspline}. Notably, we found that most motions could be represented with relatively few control points and that the number of required control points scaled only with the complexity of the motion.

\section{Parameter Values}

Tables \ref{table:handparams}, \ref{table:lambdavals}, and \ref{table:optlambdas} tabulate the parameters of our various hand models, transfer coefficients, and optimization weighting coefficients respectively. The MANO hand does not contain a skeleton and was instead animated via vertex positions. Table \ref{table:controlpoints} tabulates the total number of control points used for each motion.

\begin{table}[h!]
\centering
\caption{Parameters for all hands used in experiments}
\begin{tabular}{ |>{\centering\arraybackslash}M{2.5cm}|M{1.2cm}|M{1.2cm}|M{1.5cm}| }
    \hline
    \multicolumn{4}{|c|}{Hand Paramaters} \\
    \hline
     & \# DOFs & Root Joint & \# Phalanges \\
    \hline
    MANO Hand & Unknown & Wrist & 5 \\
    \hline
    Human Hand & 54 & Forearm & 5 \\
    \hline
    Witch Hand & 54 & Forearm & 5 \\
    \hline
    Alien Hand & 42 & Forearm & 3 \\
    \hline
    Allegro Hand & 22 & Wrist & 4 \\
    \hline
    Prosthetic Hand & 26 & Wrist & 5 \\
    \hline
\end{tabular}
\label{table:handparams}
\end{table}

\begin{table}[h!]
\centering
\caption{$\lambda$ coefficients used for bulk contact transfer}
\begin{tabular}{ |>{\centering\arraybackslash}M{1.5cm}|M{1cm}|M{1cm}|M{1cm}|M{1cm}|M{1cm}| }
    \hline
    \multicolumn{6}{|c|}{Alignment Paramter Values ($\lambda_A$, $\lambda_S$)} \\
    \hline
     & Thumb & Index & Middle & Ring & Pinky \\
    \hline
    Human Hand & 1.0, 1.2 & 1.0, 1.0 & 1.0, 1.0 & 1.0, 1.0 & 1.0, 1.253 \\
    \hline
    Witch Hand & 1.15, 1.467 & 1.55, 1.5 & 1.55, 1.593 & 1.55, 1.513 & 1.55, 1.5\\
    \hline
    Alien Hand & 1.22, 1.453 & 1.27, 1.5 & 1.232, 1.573 & N/A & N/A \\
    \hline
    Allegro Hand & 1.8, 1.9 & 1.8, 1.913 & 1.8, 1.973 & 1.8, 2.0 & N/A \\
    \hline
    Prosthetic Hand & 1.33, 1.5 & 1.33, 1.4 & 1.33, 1.4 & 1.33, 1.3 & 1.33, 1.433 \\
    \hline
\end{tabular}
\label{table:lambdavals}
\end{table}

\begin{table}[h!]
\centering
\caption{Optimization weighting coefficients used in all retargets}
\begin{tabular}{ |>{\centering\arraybackslash}M{2cm}|M{2cm}| }
    \hline
    \multicolumn{2}{|c|}{Optimization Weighting Coefficients} \\
    \hline
    $\lambda_m$ & 1.0 \\
    \hline
    $\lambda_{cd}$ & 1.0 \\
    \hline
    $\lambda_{cn}$ & 1.0 \\
    \hline
    $\lambda_c$ & 1.0 \\
    \hline
    $\lambda_t$ & 1.0 \\
    \hline
    $\lambda_j$ & 50.0 \\
    \hline
\end{tabular}
\label{table:optlambdas}
\end{table}

\begin{table}[h!]
\centering
\caption{Control point counts per motion. Motions all run at 120 FPS.}
\begin{tabular}{ |>{\centering\arraybackslash}M{2cm}|M{2cm}| }
    \hline
    \multicolumn{2}{|c|}{Control Points Used / Total Motion Frames} \\
    \hline
    Apple / Potato & 60 / 703\\
    \hline
    Doorknob & 40 / 1040\\
    \hline
    Flashlight & 80 / 1040\\
    \hline
    Hammer & 90 / 768\\
    \hline
    Phone & 120 / 1145\\
    \hline
    Stapler & 70 / 877\\
    \hline
\end{tabular}
\label{table:controlpoints}
\end{table}

We set $\mathcal{E}_{acc}$ to 500 $\frac{degrees}{s^2}$ for all angular motion and 500 $\frac{cm}{s^2}$ for all linear motion. We cap the number of iterations at 20. The same bounds were applied to all DOFs across all manipulators indiscriminately. Note that the units of linear motion are arbitrary and vary depending on the scene scale. While these thresholds are high, we found that tighter bounds were not necessary since the provided estimates were only data points for spline fits. Brief and intermittent noise from remaining large jumps in acceleration were automatically smoothed out due to our splines consisting of relatively few control points ($<$ 12\% of total frame counts).

\section{Additional Experiments}

\subsection{DOF Impact Visualization}

Because our method does not make any assumptions about the kinematics of the system, it is possible to modify the underlying skeleton and recompute the entire motion sequence without adjusting any additional information. 

A unique application of such flexibility is the ability to visualize the impact of individual DOFs for entire trajectories. In studio workflows, this capability can enable rigging artists to gain insight into whether or not introducing extra complexity is necessary for downstream animation. In robotics, the same capability can help identify which DOFs are necessary for a particular task set and which ones can be removed in favor of under-actuation. This functionality is especially valuable for rapid prototyping of new hand designs or selecting the best hand for a particular task set \cite{bauer2022iterative}.

\begin{figure}
\centering
\includegraphics[width=1.0\linewidth]{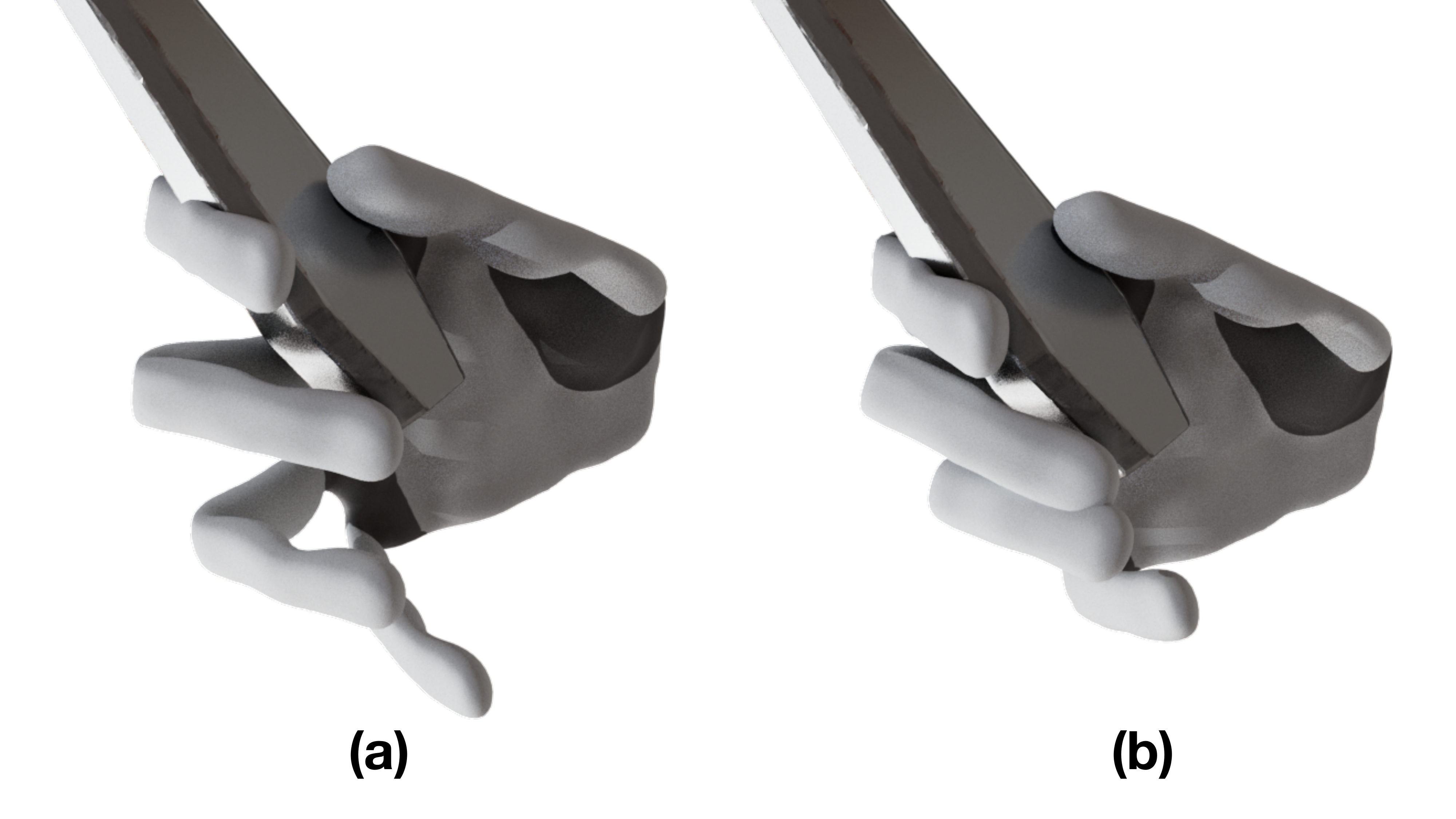}
\vspace*{-0.2in}
\caption{Frame comparison of the prosthetic with (a) one DOF and (b) two DOFs at the proximal phalange joint.}
\label{fig:prosthesisdofanalysis}
\end{figure}

We illustrate this capability in a case study of the prosthetic hand. We start by assuming one DOF at the knuckle joint of each finger controlling flexion and extension, and preform a comparison against a modified prototype which contains a second DOF at each knuckle joint controlling adduction and abduction. Figure \ref{fig:prosthesisdofanalysis} presents a representative frame of noticeable difference in intermediate solutions during a hammer manipulation.

Under the same initial configuration, optimization settings, and pipeline processing conditions, the 1-DOF knuckle variant generates intermediate solutions containing unnatural and physically implausible hammer grips. In comparison, the 2-DOF knuckle variant provides a much more natural-looking and physically believable grip. These observations suggest that the 1-DOF variant does not possess the dexterity required to manipulate the tool in an anthropomorphic manner, and that including a second motor per finger might be worthwhile. Our method can thus be used to evaluate task-centric dexterity mid-manipulation, which would otherwise be challenging to do in a manual fashion.

\section{Ablation Studies}

\subsubsection{No Root Pre-Conditioning}

\begin{figure}
\centering
\includegraphics[width=1.0\linewidth]{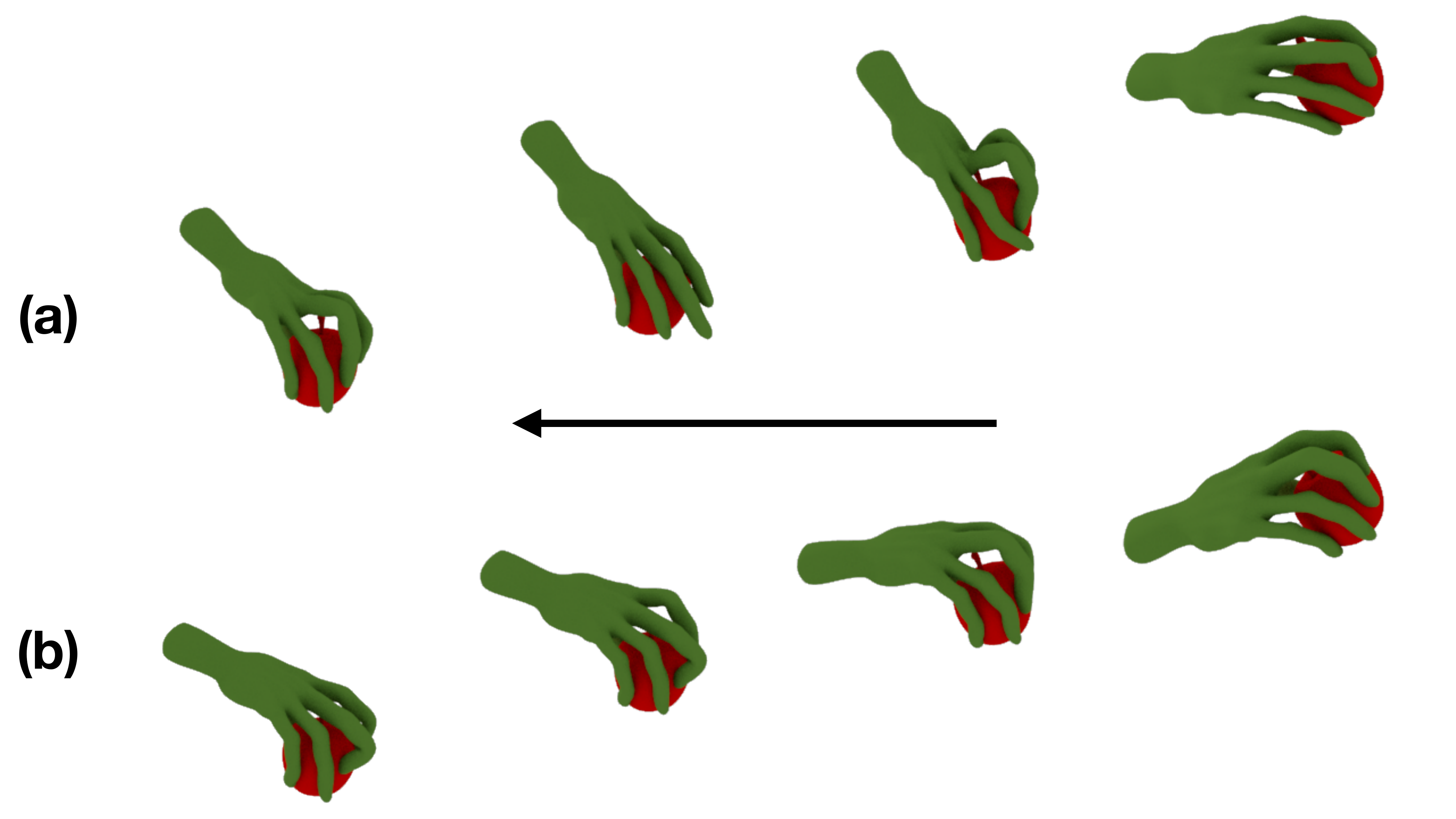}
\vspace*{-0.2in}
\caption{Example (a) without and (b) with root pre-conditioning}
\label{fig:rootablation}
\end{figure}

We examine the necessity of computing the optimal root transforms per frame. Figure \ref{fig:rootablation} illustrates the outcome for a representative sequence of motion frames.   We found results to be similar regardless of the hand, object, or motion. In summary, omission of solving for the root during an initial pass creates instances of root ``drift", in which locally optimal solutions in subsequent frames are found by moving other joints in the kinematic chain, resulting in solutions which only rotate the root joint once the accumulated marker error buildup reaches a sufficiently high level to cause the root to suddenly snap into a new optimal position with a large jump in translation. We found this issue occurred consistently among all of our hands and argue that it is because the root transformation has a disproportionately large influence over the gradients used in the solver.  Unlike other joints, a single transformation of the root joint impacts every single contact and marker point on the hand. While this is beneficial for moving the hand from a distant location to close to the object, when the hand is close to the object, the root joint eventually settles at an ``equilibrium" position (e.g. a position at which any additional movement negatively impacts the objective). Until marker error buildup becomes substantial, locally optimal solutions will instead move other joints, resulting in an undesirable solution. Solving for the root transformation exclusively directly combats this problem, providing no other way to minimize the objective through other DOFs.

\subsubsection{No Contact Data}

We perform a comparison against a pure marker tracking pipeline to answer a simple question: is contact information actually necessary for these tasks? 

\begin{figure}
\centering
\includegraphics[width=1.0\linewidth]{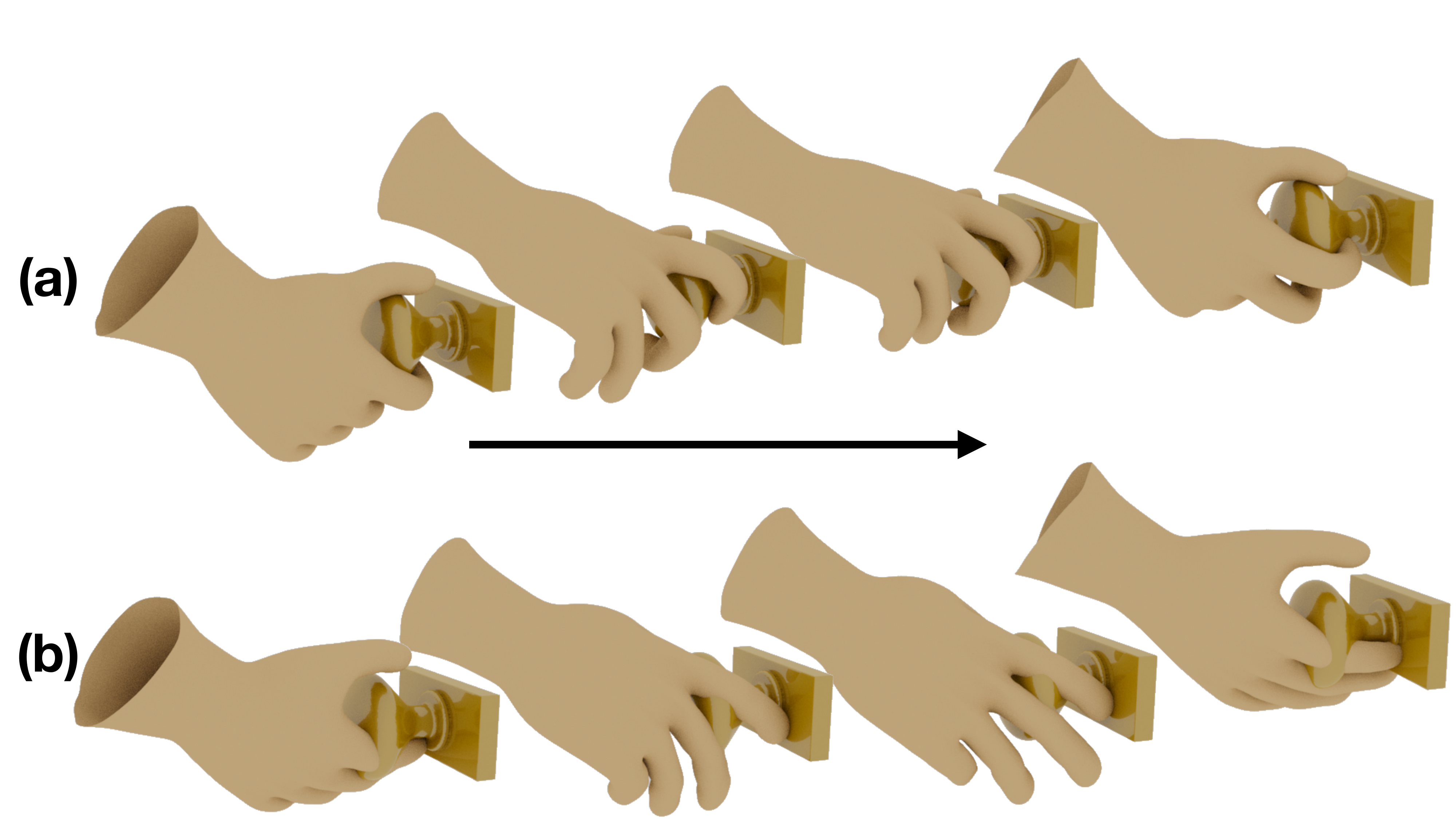}
\vspace*{-0.2in}
\caption{Sample motion frames (a) with and (b) without contact information during a doorknob manipulation.}
\label{fig:markercomparison}
\end{figure}

We select the alternate human hand for simplicity, keep both sets of virtual markers consistent, and pre-process both demonstrations with rigid wrist optimization. We then compute initial frame solutions for one motion using the original choice of parameters and another by setting $\lambda_c = 0$ to remove all contact information from objective computations. We examine results after per-frame full pose estimation since large divergences in behavior are already apparent. Figure \ref{fig:markercomparison} provides several comparison frames.

Notably, the hand lacking contact information fails to conform to the doorknob at numerous points of the manipulation, resulting in a motion estimate that is extremely poor at the outset. We note that this failure in motion estimation was compounded in the case of hands which were less human-like where marker estimates are even less reliable. The dramatic difference in motion reconstruction, even at a coarse level, clearly indicates that marker tracking information alone is simply not sufficient for motions involving rich object interaction --- contact information is essential.

\end{document}